\documentclass[
 amsmath,amssymb,
 aps, prc, showkeys
]{revtex4-2}

\usepackage{graphicx}
\usepackage{dcolumn}
\usepackage{bm}
\usepackage{hyperref}
\usepackage{float}
\usepackage[mathlines]{lineno}

\begin{document}


\title{Uncertainty band evaluation of optical potentials and differential cross-sections. Application to $^8$Li + $^{58}$Ni elastic scattering.}


\author{O.~C.~B.~Santos}
\affiliation{Departamento de F\'{\i}sica Nuclear, Instituto de F\'{\i}sica, Universidade de S\~ao Paulo, 05508-090 S\~ao Paulo, Brasil}

\author{J.~G\'omez-Camacho}%
\email{gomez@us.es}
\affiliation{Departamento de FAMN, Universidad de Sevilla, Apartado 1065, 41080 Sevilla, Spain.\\
            Centro Nacional de Aceleradores (U. Sevilla, J. Andalucía, CSIC), Tomás Alva Edison, 7, 41092 Sevilla, Spain}


\date{\today}

\begin{abstract}
A  statistical method is presented to evaluate the uncertainty bands in the optical nucleus-nucleus  potential and in differential cross sections.  The starting point is the least square fit of a set of experimental values of elastic differential cross sections, varying the relevant optical potential parameters. This is done using standard   $\chi^2$ minimization codes, that  provide the covariance matrix of the parameters. A maximum likelihood exploration of the $\chi^2$ surface in parameter space allows to determine the covariance matrix of the parameters associated to a contour of a given $\chi^2$ value. Bayes theorem allows to assign probabilities (p-values)  to the regions in parameter space, characterized by $\chi^2$ contours.  The  method allows to obtain uncertainty bands of an arbitrary observables associated to a given p-value using two approaches. The general approach determines the extremes of the observables calculated in the region of parameter space associated to that p-value. This requires an adequate sampling of parameter space, and explicit calculations of the observables on all sampling points. The simplified approach considers uncertainty propagation of the observable in terms of the optical model parameters. This involves the least-square covariance matrix, given by $\chi^2$ minimization codes,  and analytically calculated enhancement  factors for each p-value.  The method, in the general and simplified approaches,  is applied to recent measurements of the elastic differential cross sections of  $^8$Li + $^{58}$Ni. $1\sigma$ and $2\sigma$ uncertainty bands are obtained for the optical potentials as a function of the distance, and the differential cross sections as a function of the angle. The general and simplified approaches are very similar in this case. The application of the procedure to determine uncertainty bands of complex scattering calculations is discussed.
\end{abstract}


\maketitle

\section{introduction}
\label{sec:introduction}
Nuclear physics recently had important advances associated to new experimental facilities, in particular those involving radioactive beams \cite{Nupecc2024}. Structure properties of a variety of new nuclei have been determined. In many cases, the determination of  structure properties relies on the comparison of experimentally determined  cross section measurements, with the results of reaction calculations. For example, the determination of spectroscopic factors require the measurement of relevant nucleon transfer cross sections, and its comparison with the relevant transfer calculations. These calculations can be performed in the Distorted Wave Born Approximations (DWBA), or in more precise formalisms such as Coupled Channels Born Approximation (CCBA), or even Coupled Reaction Channels (CRC) \cite{DWBA, CCBA, CRC}.  The determination of electromagnetic properties, such as $B(E1)$ and $B(E2)$ transitions, require then comparison of inelastic scattering or break-up measurements, with the corresponding calculations. These reaction calculations  can be either simple semiclassic or  DWBA calculations, or more precise coupled channels (CC) or continuum-discretized coupled channels (CDCC) \cite{CC, CDCC}.

The uncertainty of the structure properties obtained by this procedure comes both from uncertainty of the experimental cross section measurements and from the uncertainty in the theoretical reaction calculations.  In recent years, the experimental nuclear physics community has made a huge effort to carry out experiments in nuclei which are further away  from the stability, improving in the acceleration and in the detection techniques. The experimental uncertainties are carefully controlled, and made as small as possible, within the experimental limitations, such as the beam time allocated. Thus, it is fair to request that the theoretical nuclear physics community involved in nuclear reactions dedicates efforts to provide the uncertainty of the calculated theoretical cross sections.

The  main sources of theoretical uncertainties in cross section calculations are known, and some of them have been outlined in ref \cite{Nunes2017}. The state of the art in nuclear reaction theory is well described in the classic book of Satchler \cite{satchler}, and on this basis, a hierarchy of theoretical treatments can be established. It is known, for example, that quantum calculations, such as DWBA, are more precise than semi-classical or eikonal calculations for the evaluation of inelastic and transfer cross sections. It is known that CC calculations are more precise than DWBA calculations, for the evaluation of inelastic scattering, as well as to describe the effects of the coupling in the elastic channel. It is known that CDCC calculations, including the effect of coupling to the continuum, are more precise than CC calculations, that treat resonances as bound states.  It is known that CRC calculations are more precise than DWBA calculations for the calculation of transfer cross sections. It is known that Faddeev Calculations, when feasible, are more accurate than CRC or DWBA calculations for rearrangement collisions.  To our knowledge, there are no systematic, general calculations that allow to quantify the uncertainty that one has when using a simple formalisms, such as the eikonal approximation for neutron removal, DWBA for transfer, or semiclassiclal calculations for Coulomb excitation, instead of more precise formalisms. It is fair to say that, in many cases, the nuclear reactions community tries to use the best available state of the art reaction calculations for a given purpose. However, this is not enough. The best calculation, from a formal point of view, of a scattering magnitude, has uncertainties associated to the uncertainties of the ingredients used in the calculations. 

Possibly, the largest uncertainty in any scattering calculations, where nuclear potentials play a role, arises from the uncertainties in the optical potentials used in the calculations. Optical potentials are essential ingredient to evaluate any scattering magnitudes in nucleon-nucleus and nucleus-nucleus collisions. They appear in Optical Model, DWBA, CC, CDCC, CRC and even Faddeev calculations. 
Optical potentials are usually described by phenomenological real and imaginary potentials, each of them described by three parameters: The depths $V, W$, the reduced radii $r_V, r_W$, and the diffuseness $a_V, a_W$.  Indeed, in some cases, microscopic models can be used to describe the real part of the optical potentials, but the imaginary potential remains phenomenological. In other cases, complex nuclei can be described in terms of core and valence particles, but the optical potentials between core, valence and target remain neccessary. In principle, it is expected that an accurate measurement of the differential elastic cross section, at a given scattering energy, should allow to determine the optical potentials, for each scattering energy. Although many parameterizations are available in the literature for the optical potentials for different pairs of projectile and targets, there is not a systematic assessment of the uncertainties of these parameters.

Nunes and collaborators  \cite{Nunes2017, Nunes2019, Nunes2022, Nunes2024}, have faced the challenge of evaluating uncertainties for a variety of nuclear reactions.  In their pioneering work,\cite{Nunes2017, Nunes2019}, they deal with the uncertainty of the optical model parameters. They compare the {\em frequentist} approach, which is used in the standard $\chi^2$ fitting codes, with a bayesian method, based on Markov Chain Monte Carlo  (MCMC). It is found that the {\em frequentist} uncertainties are significantly smaller than those in Bayesian approach. Also, the strong correlations between the parameters found in the {\em frequentist} approach do not seem to appear in the Bayesian treatment. They have applied this formalism to the evaluation of break-up reactions \cite{Nunes2022} and to charge exchange reactions \cite{Nunes2024}. A very important aspect of these works is that they provide uncertainty bands for the theoretical calculations.
These bands indicate the expected range of the theoretical calculations, for a given confidence level.

In section \ref{sec:uncertaintyevaluation} we present the method, which is motivated by the previously cited works  \cite{Nunes2017, Nunes2019}. For a better understanding of the discrepancies between the  uncertainties in standard $\chi^2$ fitting codes and in the Bayesian approach, we avoid the term  {\em frequentist}, and follow the terminology used in \cite{cowan98}. We start with the least-square approach, implicit in most $\chi^2$ fitting codes. Then we consider the closely related maximum likelihood method, which requires to explore the parameter space beyond the vicinity of the  $\chi^2$ minimum. Finally, we introduce Bayes theorem, which allows to assign p-values to different regions in parameter space, and is essential to introduce p-values in the uncertainty evaluation. To make the discussion more clear, we avoid the use of Monte Carlo Methods, and consider instead a wide range of sampling points in the parameter space.
We focus on the evaluation of uncertainty bands of optical potentials as a function of the radius, and of differential cross section calculations as a function of the angle.   We obtain  the uncertainty of scattering calculations using two approaches.
The general approach is done by explicit calculations in all the sampling points of a given parameter region characterized by a p-value.  The approximate approach makes use of uncertainty propagation,  taking as input the least-square covariance matrix of the parameters, provided by standard $\chi^2$ codes, and analytic enhancement factors arising from the Bayesian treatment. 

In section \ref{sec:application}
we apply our method to experimental data for the elastic scattering of $^8$Li+$^{58}$Ni at several energies just above the Coulomb Barrier, considering the experimental uncertainties. This is a heavy ion collision, that differs from the nucleon-nucleus systems considered in ref \cite{Nunes2019}. In our case,  the elastic differential cross sections are mostly sensitive to the tail of the potential. So, rather than determining independently the six parameters, one may fix $r_V, r_W$, and determine the other four parameters $V, W, a_V, a_W$ to reproduce the elastic differential cross sections. This is the well known Igo ambiguity \cite{igo}. As a result of our procedure, we evaluate the uncertainties of the optical model parameters, from which we derive uncertainty bands for the optical potential as a function of the radius, and uncertainty bands for the differential cross sections as a function of the angle. These uncertainty bands  are associated to $1\sigma$ and $2\sigma$ statistical significance, corresponding respectively to p-values of 31.8\% and 4.5\%. 
In section \ref{sec:summary} we summarize, discuss and present the conclusions of this work. Appendix \ref{ap:leastSquare} collects the expressions linking least square, maximum likelihood and Bayesian treatments.

\section{Method for uncertainty band evaluation}
\label{sec:uncertaintyevaluation}

\subsection{Notation}

Let us introduce the notation that will be used through this paper. We will follow, as much as possible, the conventions used in \cite{cowan98}. We consider a set of observable quantities  $y_i$, corresponding typically to elastic differential cross sections, and $i$ labels the scattering angle and scattering energy. We consider an experimental  measurement of these quantities, whose results are given by values $\hat{y}_i$, and uncertainties $\sigma_i$. The index $i$ (and eventually $i'$) characterizes different experimental conditions, and it ranges from $i=1$ to $i=N$, where $N$ is the number of experimental points. We consider that each observable quantity considered $y_i$ is described by a normal probability distribution with mean $\hat{y}_i$ and standard deviation $\sigma_i$, and that there are no correlations between different measurements $i$. Thus, in probabilistic terms, if the observable $y_i$ is measured several times, the obtained results  has a probability  68.2 \% of being in the range $(\hat{y}_i - \sigma_i, \hat{y}_i + \sigma_i)$, and a probability 31,8 \% of being outside.   ${y}$ indicates the N-dimensional vector $(y_1, \dots, y_N)$ of the observable quantities.  $\hat{y}$ indicates the N-dimensional vector $(\hat y_1, \dots, \hat y_N)$ of the experimental results.

We will use model descriptions of $y_i$, typically optical model calculations, which depend on series of parameters $a_j$, typically optical model parameters. The index $j$ (and eventually $k,l$) labels the parameters, and it ranges from $j=1$ to $j=M$, where $M$ is the number of parameters.   We will refer to a parameter vector ${a}$, which has $M$ components that are the parameters $a_j$, and so it lives in an $M$-dimensional parameter space. The model description provides a theoretical prediction of each one of the observable quantities characterized the index $i$, which depends on the parameter vector $a$. This prediction is labelled as $y_i(a)$. The overall agreement of the model, considering the parameter vector $a$,  with the experimental data is measured by the quantity $\chi^2(a) = \sum_i (y_i(a)-\hat{y}_i)^2/\sigma_i^2$. The function  $\chi^2(a)$, which is positive definite, should have an absolute minimum  $\chi^2_m $ corresponding to the parameter vector $\hat{a}$, so that  $\chi^2_m = \chi^2(\hat a) $. The components $\hat{a}_j$ of the parameter vector $\hat a$, are obtained from a $\chi^2$ minimization procedure. The first derivatives of $\chi^2(a)$ evaluated at $\hat a$ vanish.The second derivatives of $\chi^2(a)$ evaluated at $\hat a$ define the Hessian matrix $H_{jk}$

We use the notation $\chi^2(a)$, with round brackets, to indicate that $\chi^2(a)$ is a function of the model parameters $a$, that has a minimum $\chi^2_m$, and a Hessian matrix $H_{jk}$, determined by the second derivatives. 
The notation  $\chi^2[p]$, with square brackets, indicate  a value of $\chi^2$, which is associated to a given probability (p-value)  $p$ in the  probability distribution of the parameter space that will be derived using Bayes theorem. Closely related to the p-value is the statistical significance, which is indicated by multiples of $\sigma$. A p-value $p$ corresponds to a statistical significance $n\sigma$ when $p$ is the probability in the normal distribution of the value being outside of $\pm n \sigma$.
A p-value $p=0.318$ corresponds to a $1\sigma$ statistical significance.
An equivalent notation to $\chi^2[p]$ is $\chi^2[n\sigma]$,  So, $\chi^2[p=0.318] \equiv \chi^2[1\sigma]$. The values $b$ used  in the maximum likelihood approach are related to a given confidence level through  $b = \chi^2[p]$. They define a region in parameter space given by the values of $a$ fulfilling $\chi^2(a) \le b$. 

\subsection{Uncertainty evaluation of the parameters}

In order to describe the uncertainties of the parameters, we consider three approaches, with increasing complexity and generality:
In the least square approach (LS),  one considers small variations of the parameter vector $a$ with respect to the minimum $\hat a$. The possible deviation of the parameters $a_j$ with respect to the minimum values  $\hat a_j$,  are given by a linear combination of the possible deviation of the experimental values with respect to the nominal values $\hat y_i$. 
The covariance matrix of the parameters $cov_{LS}(a_j, a_k)$ can be obtained from error propagation from the uncertainties of the data, and are determined by the Hessian matrix $H_{kj}$. The expression of the covariance matrix, consistent with \cite{cowan98} and derived in the appendix \ref{ap:leastSquare}, is

\begin{equation}
cov_{LS}(a_k, a_j) =  2 H^{-1}_{kj}. \label{covhess2}
\end{equation}
The standard deviation of the parameters is given by the square root of the diagonal elements of the covariance matrix 
\begin{equation}
    \sigma(a_j) = \sqrt{cov_{LS}(a_j, a_j)}.
\end{equation}
The correlation matrix is obtained from the covariance matrix, divided by the product of the standard deviations 
\begin{equation}
    corr_{LS}(a_j, a_k) = {cov_{LS}(a_j, a_k) \over \sigma(a_j)\sigma(a_k)}.
\end{equation}
It should be stressed that the expression of the covariance \ref{covhess2} does not take into account any probability information related to the quality of the fit. 
In deriving  equation~(\ref{covhess}) we did not consider at all the actual value of $\chi^2_m$. We just imposed the conditions \ref{partchi2}, which define an implicit relation of the parameters and the experimental values $\hat y_i$, and carried out uncertainty propagation. The LS approach, which explores only the vicinity of the minimum, can provide a covariance matrix of the parameters  $ cov_{LS}(a_j, a_k)   $, but does not allow to evaluate uncertainty bands of given statistical significance.


Other approach to describe the uncertainties  is the maximum likelihood method. In it, we consider the parameter values $a$ are characterized by a likelihood function ${\cal L}(\chi^2)$, which depends on $\chi^2(a)$. The points in parameter space characterized by $\chi^2(a) = b$ define a contour in the M-dimensional parameter space with the same likelihood ${\cal L}(b)$.  When $b= \chi^2_m$, the contour reduces to a single point $\hat a$, and as $b$ increases, the contour becomes larger, and the relevant points $a$ differ more from $\hat a$. All these points  have associated a covariance matrix  $cov[b](a_j, a_k)$, that depends parametrically on the parameter $b$. The expressions are derived in the appendix. In particular, when the $\chi^2(a)$ surface is approximately parabolic, it becomes proportional to the least-square covariance matrix
\begin{equation}
   cov[b](a_j, a_k)   = f[b]^2 cov_{LS}(a_j, a_k)   \label{covbsmooth1}
\end{equation}
where $f[b] = \sqrt{(b-\chi^2_m)/M}$ is an enhancement factor. Its values, for some cases of interest are given in Table \ref{tablechi2}.

The maximum likelihood method allows to explore the full parameter space, and so it can provide covariance matrices of the parameters $  cov[b](a_j, a_k)  $ for regions in the parameter space characterized by $\chi^2(a) = b$, which can be arbitrarily far away from the minimum  $\hat a$. However,
the likelihood function ${\cal L}(\chi^2)$ is not a probability distribution, so, unless some additional considerations are done, one cannot evaluate uncertainty bands of given statistical significance.

Finally, we consider the Bayesian approach. This approach takes into account the probability distribution associated to the $\chi^2$ values. 
The Bayesian approach allows to assign p-values to the parameter regions limited by the contours $\chi^2(a) = b$. As it is derived in the appendix, the
Bayesian posterior probability that the parameters $a$ are outside a region  $\Gamma[b]$  defined by  $ \chi^2(a) \ge b \ge\chi^2_m$, is given by
\begin{equation}
        p(\Gamma[b]|H) =  {P(b, L)  \over   P(\chi^2_m, L)} , \label{Bayesp}
\end{equation}
where $ P(b, L)   $ is the cumulative $\chi^2$ probability distribution for $L$ degrees of freedom, described in the appendix.
$   p(\Gamma[b]|H)  $ corresponds to the p-value, that states the probability that the true values of the parameters $a$ are outside the region $\Gamma[b]$.  
Thus, the Bayesian approach allows to obtain the probabilities associated to $b= \chi^2(a)$ values, which were missing in the maximum  likelihood method.
For a given statistical significance in terms of $n\sigma$, associated the corresponding p-value, one can obtain the value of $b$ using eq. \ref{Bayesp}. This allows to determine the region $\Gamma[b]$ in parameter space, as well as the enhancement factor $f[b]$ defined previously.

Using the simplified expression \ref{covbsmooth}, and implementing analytical properties of the $\chi^2$ distribution functions, we get
\begin{eqnarray}
cov_B(a_j, a_k) &=&  F^2_B \;   cov_{LS}(a_j, a_k) \label{covbayesiansmooth}\\
F^2_B  &=&   {L \; P(\chi^2_m, L+2) - \chi^2_m \; P(\chi_m^2, L) \over M \; P(\chi_m^2, L)}
\end{eqnarray}
This expression indicates that the Bayesian approach produces a systematic enhancement of the covariance matrices, as compared to the least square approach, given by $F_B$.  

\begin{table}[h]
\caption{ \label{tablechi2} p-values, $\chi^2$ values and enhancement factors for various statistical significancies,  for a case with $L=48$, $\chi^2_m= 47,32$, $M=4$ . The Bayesian enhancement factor $F_B$ is 1.45}
\begin{tabular}{cccccc}
Significance  &  p-value      & $b$ & $f[b]$  \\ \hline
$0 \sigma$        &  1        & 47.32               &   0     \\
$0.5  \sigma$     &  0.61708  & 52,35               &	1.12  \\
$1 \sigma$	      &  0,31731  &	57,72               &	1,61  \\
$1.5  \sigma$     &  0,13361  &	63,44               &	2,01  \\
$2  \sigma$       &  0,04550  &	69,52               &	2,35  \\
$2.5 \sigma$      &  0,01242  & 75,95               &	2,68  \\
$3\sigma$         &  0,0027   & 82,76               &	2,97  \\\hline \hline
    \end{tabular}
\end{table}

\subsection{Uncertainty evaluation of calculated quantities}

Let us consider a generic quantity $O$, which can be calculated as a function of the parameters $O(a)$. The maximum likelihood value of $O$, which coincides with the least square estimate value, and with the value with the largest Bayesian  probability density,  corresponds to the calculations using the parameter vector $\hat a$ that produces the minimum $\chi^2_m = \chi^2(\hat a)$. This calculation will be named as $\widehat O = O(\hat a)$.
To evaluate the uncertainties of this magnitude we use the general approach and the simplified approach.

In the general approach, we obtain all the values that the quantity $O(a)$ can take, considering that the parameter vector $a$ belongs to a region in parameter space $\Gamma[b]$ which is characterized by a given statistical significance $n\sigma$, which is equivalent to a given p-value $p$. As shown in Table \ref{tablechi2}, the value of the parameter $b=\chi^2[p]$ which defines the region $\Gamma[b]$ in parameter space defined by $\chi^2_m \le \chi^2(a) \le b$ associated to this p-value  can be determined. Then, the uncertainty band of the quantity $O$, associated to the p-value $p$ is just the range of values that $O(a)$ takes in that region. A sufficient sampling of the parameter space in the region $\Gamma[b]$ should be done, and the maximum and minimum value of $O(a)$, $O_x[b]$ and $O_m[b]$, can be determined. The uncertainty range will in general not be symmetric, as $ O_x[b]- \widehat O \ne \widehat O - O_m[b]  $. This procedure is straightforward although indeed it is computationally demanding, as many calculations of $O(a)$ are required.

In the simplified approach, we assume that $O(a)$ is a smooth function of the parameter vector $a$, that can be Taylor-expanded around the minimum $\hat a$.  The least square standard deviation of the observable $O$ can be obtained using uncertainty  propagation, 
\begin{eqnarray}
\sigma_{LS}(O) &\simeq & \sqrt{\sum_{kj} O'_k  cov_{LS}(a^e_k, a^e_j) O'_j}  \label{eq.sigmaLSO.deriv}\\
O'_j & = & \left. {\partial O(a) \over \partial a_j} \right|_{a = \hat a}
\end{eqnarray} 
Notice that the derivatives $O'_j$ of the calculated quantity with respect to the parameters can be approximated  numerically, by performing the differences of the calculations of the observables $O(a)$ where the parameters $a_j$ are set as $\hat a_j \pm \sigma(a_j)$. This allows to express the standard deviation on terms of the correlation matrix and the differences $\delta O_j $
\begin{eqnarray}
\sigma_{LS}(O) &\simeq& \sqrt{\sum_{kj} \delta O_k  \; corr_{LS}(a_k, a_j) \; \delta O_j} \label{eq.sigmaLSO.delta}\\
\delta O_j & = &  {O(a_j+\sigma(a_j))-O(a_j - \sigma(a_j)) \over 2}
\end{eqnarray} 

Instead of the standard deviation $ \sigma_{LS}(O) $ , we want to evaluate the uncertainty range associated to a given p-value $p$, which correspond to the points in the region $\Gamma[b]$ of the parameter space.  If $O(a)$ is a smooth function of $a$, the maximum and minimum values of $O(a)$, $O_x[b], O_m[b]$ should occur for values of $a$ that are in the contour $\chi^2(a)=b$. In particular, they will be in the directions in parameter space determined by the derivatives $O'_k$. The range of values, symmetric in this approach, is given by 
$\sigma[b](O) = O_x[b]-\widehat O = \widehat O- O_m[b] $, where
\begin{equation}
\sigma[b](O) = \sqrt{\sum_{kj}O'_k \; cov[b](a^e_k, a^e_j) \; O'_j} = f[b] \sigma_{LS} (O)
\label{eq.sigmabO}  
\end{equation}
$f[b]$ is the adequate enhancement factor for the value $b=\chi^2[p]$ corresponding to a p-value $p$. 
The simplified approach is very easy to implement, and it is not computationally demanding. The evaluation of the differences $ \delta O_j $ requires only to perform $2M$ calculations of $O(a)$, and the correlation matrix is provided by the $\chi^2$ fit code.
The comparison of the simplified approach with the more demanding general approach provides an assessment of the adequacy of the method.

\section{Application to \texorpdfstring{\boldmath${}^{8}$Li + ${}^{58}$Ni\unboldmath}{8Li + 58Ni} elastic scattering}
\label{sec:application}

\subsection{Least square fits}
\label{sec:leastsquare}

\begin{figure}[htb]
    \centering
    \includegraphics[width=0.5\linewidth]{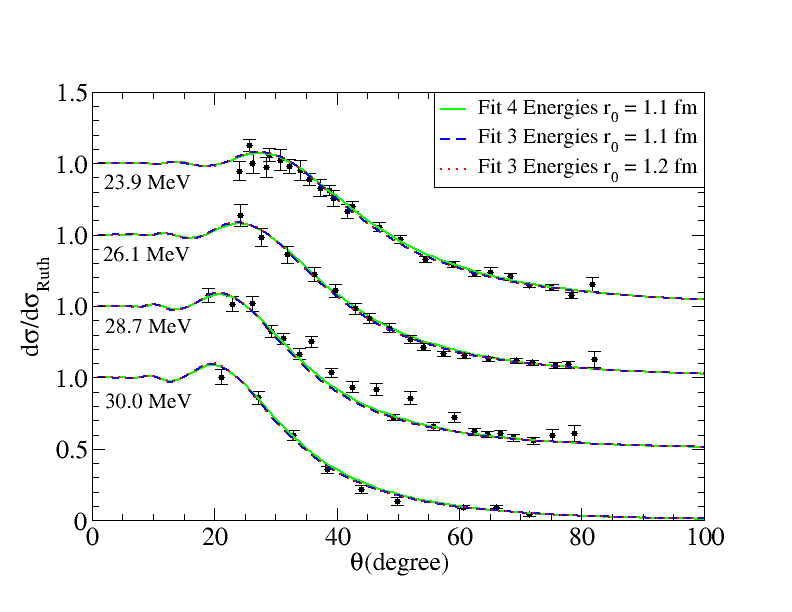}
    \caption{Elastic scattering angular distribution compared with the Woods-Saxon optical model fits, including three sets of data ($r_0= 1.1$ and $r_0= 1.2$ fm), and four sets of data $r_0= 1.1$ fm. }
    \label{fig:el}
\end{figure}

We considered elastic cross section data, from \cite{data}, at four different energies, 23.8, 26.1, 28.7 and 30.0 MeV.
The analysis done in \cite{data} indicate that the data at 28.7 MeV, has a different behaviour from the other energies. Then, we have performed fits considering three sets of data, corresponding to three energies
23.8, 26.1 and 30.0 MeV, and fits considering the four sets of data, at energies, 23.8, 26.1, 28.7 and 30.0 MeV.
In all cases, we performed 4 parameter fits for the real and imaginary potential parameters, varying $V, a_V, W, a_W$. The radial parameters $r_0$ were not varied, as they would be redundant with the depth of the potential, due to the well known Igo ambiguity \cite{igo}. Nevertheless, for the three sets of data we have performed fits with $r_0=1.1$ and $r_0=1.2$ fm to check the consistency of the results. For the four sets of data we just took $r_0=1.1$ fm.
Initially we tried to fit different potentials for different energies. However, the obtained parameters have very large uncertainties, associated to large correlations between different parameters. Also, the $\chi^2_m$ values obtained are significantly smaller than the corresponding degrees of freedom. Thus, within the experimental uncertainties considered, the elastic scattering data are not able to constrain sufficiently an energy dependent  optical potential. We have fitted the data (three or four energies),  with a single, energy independent interaction. The results are presented in Table \ref{tab:Hesse_fix_r0}.  The fits of the three sets of data are satisfactory, as the $\chi^2_m$ values are very close to the number of degrees of freedom, indicating that an energy independent optical potential can reproduce satisfactorily the differential cross section at these three energies. We also see that the quality of the fits are identical for the different $r_0$ values, as we should expect from Igo ambiguity \cite{igo}.  The fits of the four sets of data are much worse, as the $\chi^2_m$ values are about two times the number of degrees of freedom. The discrepancy is due to the contribution of the energy at 28.7 MeV, which has a different trend compared to the others.
In figure \ref{fig:el} we compare the optical model calculations with the data.
Notice the discrepancy for the energy 28.7 MeV, which appears both in the three sets calculations, that to not include these data, and in the four sets calculations, that include the data. The values and standard deviations of the potential parameters, obtained in the least square method, are shown in Table~\ref{tab:Hesse_fix_r0}. 

The first observation in all the fits is that the imaginary potential depth is considerably larger than the real potential depth. This dominance of imaginary potentials was found in the scattering of halo nuclei, such as $^6$He~\cite{sanchezbenitez}.
The uncertainties in the real potential parameters are, in relative terms, larger than those of the imaginary potential. The correlation matrix,  shown in Table~\ref{tab:corre_fix_r0}, indicate that uncertainties of the parameters are significantly correlated, although the correlation is reduced with respect to the fits of energy-dependent optical potentials. The negative correlations between $V$ and $a_V$, on one side, and between $W$ and $a_W$, on the other side, are specially significant. This indicates that, starting from the best fit values,  $V$ may be increased by its uncertainty, and $a_V$ decreased by its uncertainty, or vice versa, and the fit will be similarly good. The application of the simplified approach to evaluate uncertainties of arbitrary magnitudes just requires the standard deviation  of the parameters, given in Table \ref{tab:Hesse_fix_r0}, the correlation matrix, given in Table \ref{tab:corre_fix_r0} , and the enhancement factors   given in Table \ref{tab:sigUnce_fix_r0}. 

\begin{table}[htb]
    \centering
    \caption{Wood-Saxon potential parameters and standard deviations (in brackets), obtained from a $\chi^2$ fit at the three energies considered (3 sets), $r_0=$ 1.1 and 1.2 fm. Also presented the results for the fit including four energies (4 sets).}
    \begin{tabular}{r r r r r r r r r r}
    r$_{\mathrm{0}}$(fm) &  $V$(MeV)   & $a_V$(fm)     & $W$(MeV)   & $a_W$(fm)      & $\chi^2$  & $\chi^2/L$  &  N   &  p  & L   \\ \hline
        1.1 (3 sets)             &  8.3(45)  & 0.95(16)   & 97(35)   & 0.624(36)   & 47.34     & 0.9864      &  52  &  4  & 48  \\
        1.2  (3 sets)            &  5.8(47)  & 0.88(22)   & 47(17)   & 0.597(57)   & 47.53     & 0.9902      &  52  &  4  &  48 \\ 
             1.1  (4 sets)            &  4.0(25)  & 1.08(24)   & 63(17)   & 0.671(39)   & 131.91     & 1.9117      &  73  &  4  & 69  \\  
        \hline \hline
    \end{tabular}
    \label{tab:Hesse_fix_r0}
\end{table}

\begin{table}[!htb]
    \centering
    \caption{Correlation matrix of the optical potential parameters obtained with fixed reduced radii, for 3 sets and 4 sets of data.}
    \begin{tabular}{ c c c c c | c c c c c}
        \multicolumn{5}{c|}{r$_{\mathrm{0}}=1.1~\mathrm{fm}$ 3 sets} &   \multicolumn{5}{c}{r$_{\mathrm{0}}=1.2~\mathrm{fm}$ 3 sets} \\
            &  $V$    & $a_V$    & $W$    & $a_W$                       &    &   $V$    &  $a_V$     &  $W$      &  $a_W$    \\
        $V$   & 1.000 &-0.930 & 0.758 &-0.621                   & $V$  &  1.000 & -0.946  &  0.911  & -0.834 \\
        $a_V$  &-0.930 & 1.000 &-0.572 & 0.319                   & $a_V$ & -0.946 &  1.000  & -0.784  &  0.622 \\
        $W$   & 0.758 &-0.572 & 1.000 &-0.882                   & $W$  &  0.911 & -0.784  &  1.000  & -0.931 \\
        $a_W$  &-0.621 & 0.319 &-0.882 & 1.000                   & $a_W$ & -0.834 &  0.622  & -0.931  &  1.000 \\ \hline 
     \multicolumn{5}{c|}{r$_{\mathrm{0}}=1.1~\mathrm{fm}$ 4 sets} &   \multicolumn{5}{c}{} \\
            &  $V$    & $a_V$    & $W$    & $a_W$                       &    &   &&&   \\
        $V$   & 1.000 &-0.899 & 0.849 &-0.775                   &&&&&  \\
        $a_V$  &-0.899 & 1.000 &-0.704 & 0.470                   &&&&&  \\
        $W$   & 0.849 &-0.704 & 1.000 &-0.895                   &&&&& \\
        $a_W$  &-0.775 & 0.470 &-0.895 & 1.000                   &&&&&  \\ \hline 
        
        \hline
    \end{tabular}
    \label{tab:corre_fix_r0}
\end{table}

\begin{table}[htb]
    \centering
    \caption{Enhancement factors for the uncertainties, for various fits.}
    \begin{tabular}{r r r r r r r}
    r$_{\mathrm{0}}$(fm) &  $b_{1\sigma}$  & $b_{2\sigma}$ & $b_{3\sigma}$  & $f[b_{1\sigma}]$  & $f[b_{2\sigma}]$ &  $f[b_{3\sigma}]$  \\ \hline
        1.1  (3 sets)            &  57.72              & 69.52             & 82.76              & 1.612          & 2.356         &    2.977       \\
        1.2   (3 sets)           &  57.85              & 69.61             & 82.84              & 1.606          & 2.349         &    2.971       \\ 
        1.1   (4 sets)           &  136.38            & 143.66             & 153.76              & 1.057          & 1.714         &    2.337       \\ 
        \hline \hline
    \end{tabular}
    \label{tab:sigUnce_fix_r0}
\end{table}

\subsection{Exploration of the \texorpdfstring{$\chi^2$}{chi2} surface as a function of the optical potential parameters}

The enhancement factors presented in Table \ref{tab:sigUnce_fix_r0}, as well as the correlation matrix \ref{tab:corre_fix_r0} were derived in the simplified approach, which assumes that the $\chi^2(a)$ surface is parabolic.  The general approach requires the evaluation of the  $\chi^2$ surface  as a function of the parameters. We present two-dimensional plots of   $\chi^2$ surface versus the six pairs of parameters that can be obtained from $V$, $a_V$, $W$, $a_W$, indicating the contour plots of the $\chi^2$ values corresponding to statistical significances of $1\sigma$, $2\sigma$ and $3\sigma$. The most prominent feature is the strong correlations that were present  $V$ and $a_V$, and also, to a lesser degree, between $W$ and $a_W$. This is fully consistent with the values in the correlation matrix  \ref{tab:corre_fix_r0}. Notice also that the  $\chi^2$ contours are not elliptical, and this indicates deviations of $\chi^2(a)$ from the parabolic behavior. So we can expect some differences between the general and the simplified approaches.

\begin{figure}
    \centering
    \includegraphics[width=0.32\linewidth]{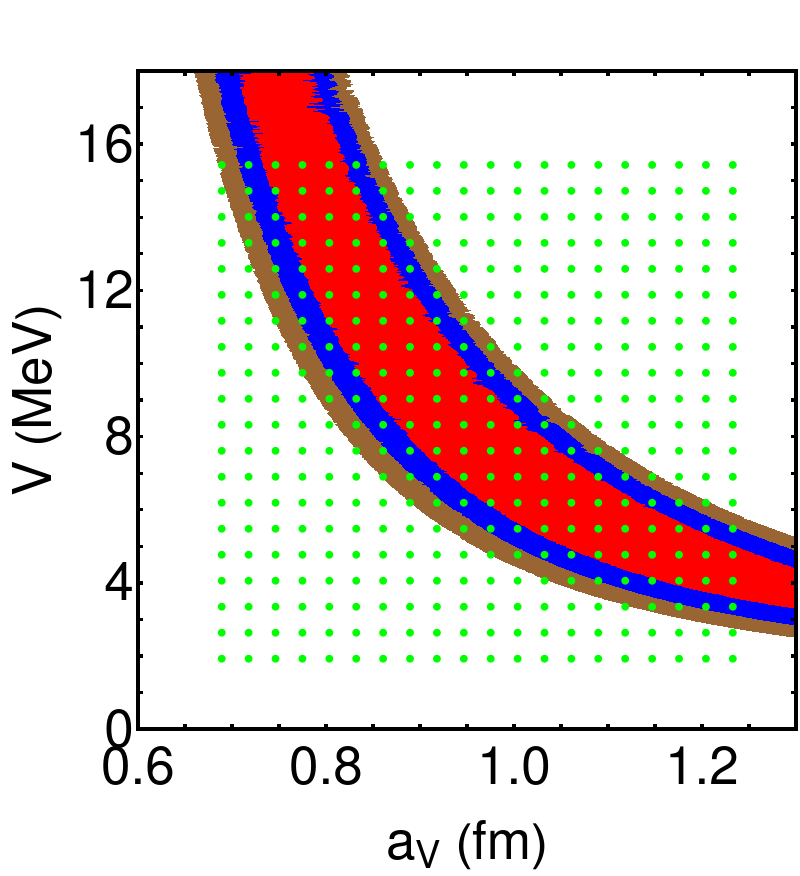}
    \includegraphics[width=0.32\linewidth]{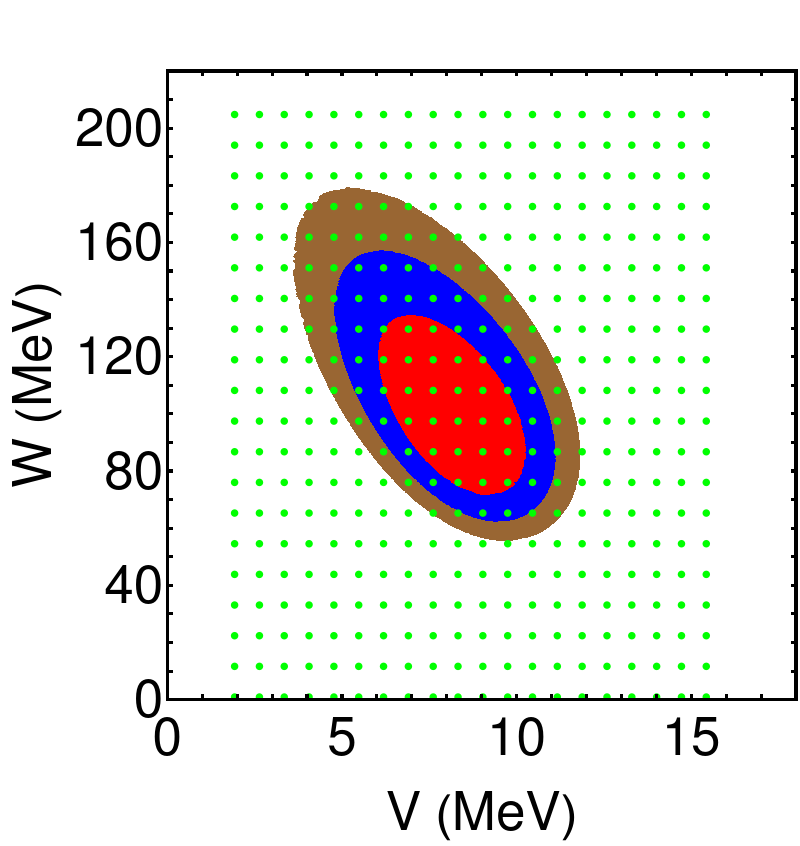}
    \includegraphics[width=0.32\linewidth]{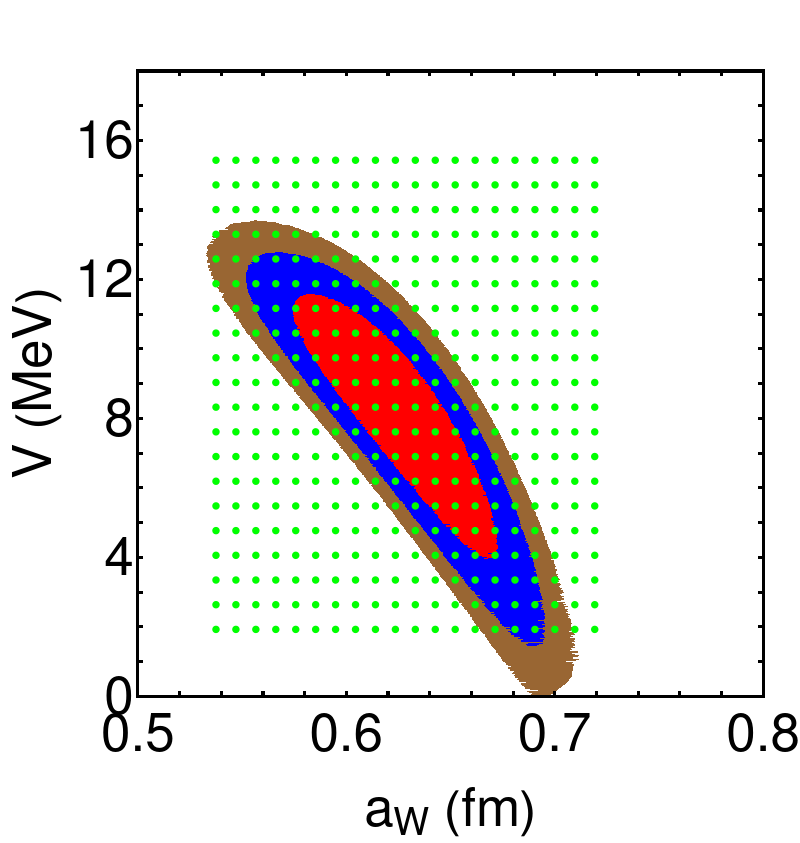}\\
    \includegraphics[width=0.32\linewidth]{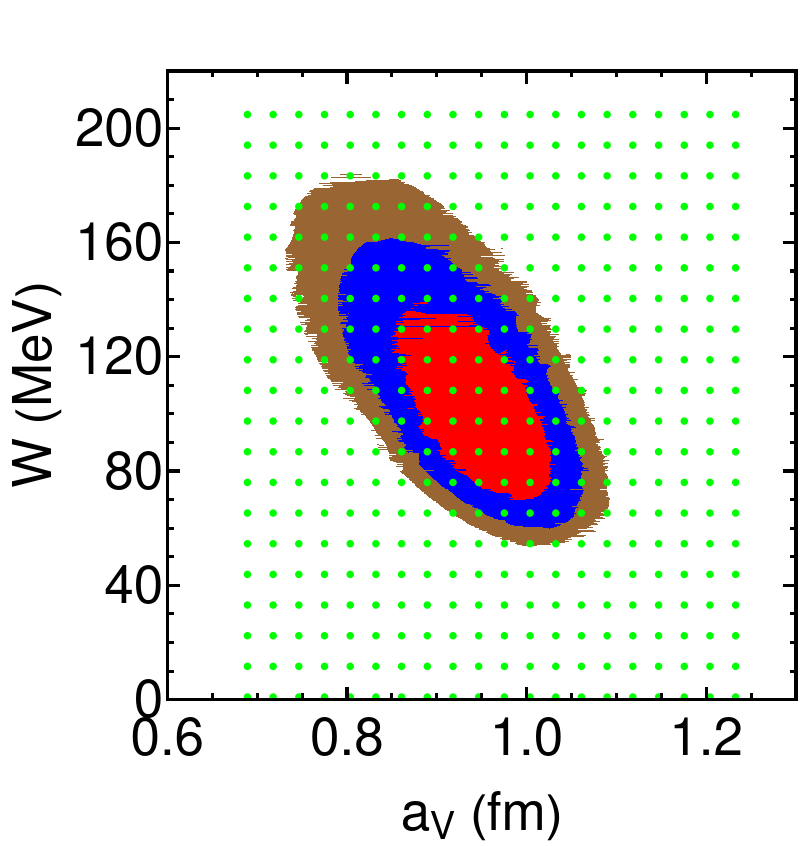}
    \includegraphics[width=0.32\linewidth]{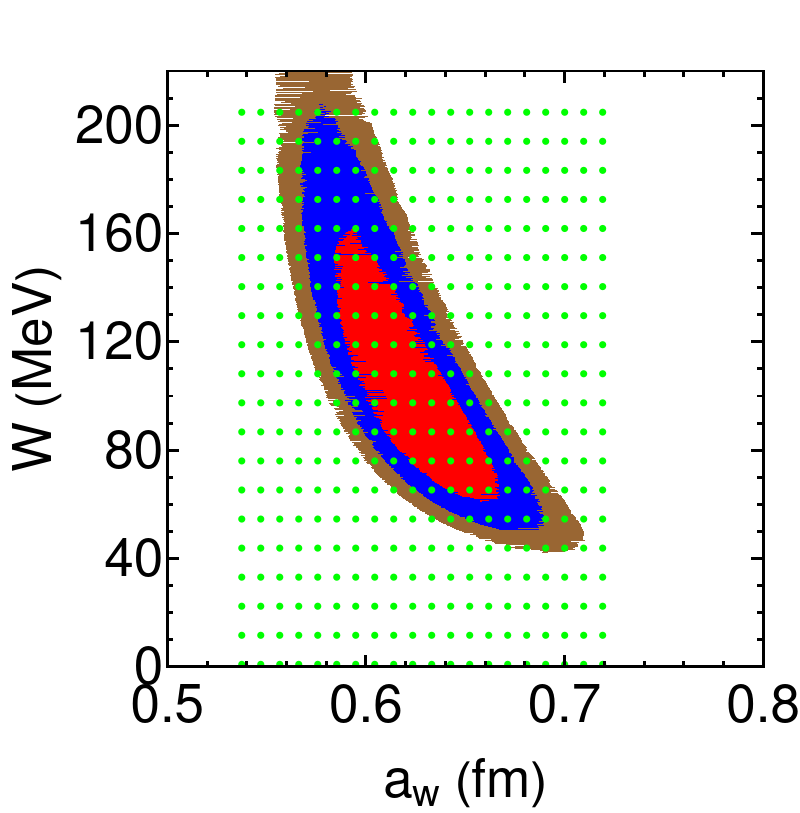}
    \includegraphics[width=0.32\linewidth]{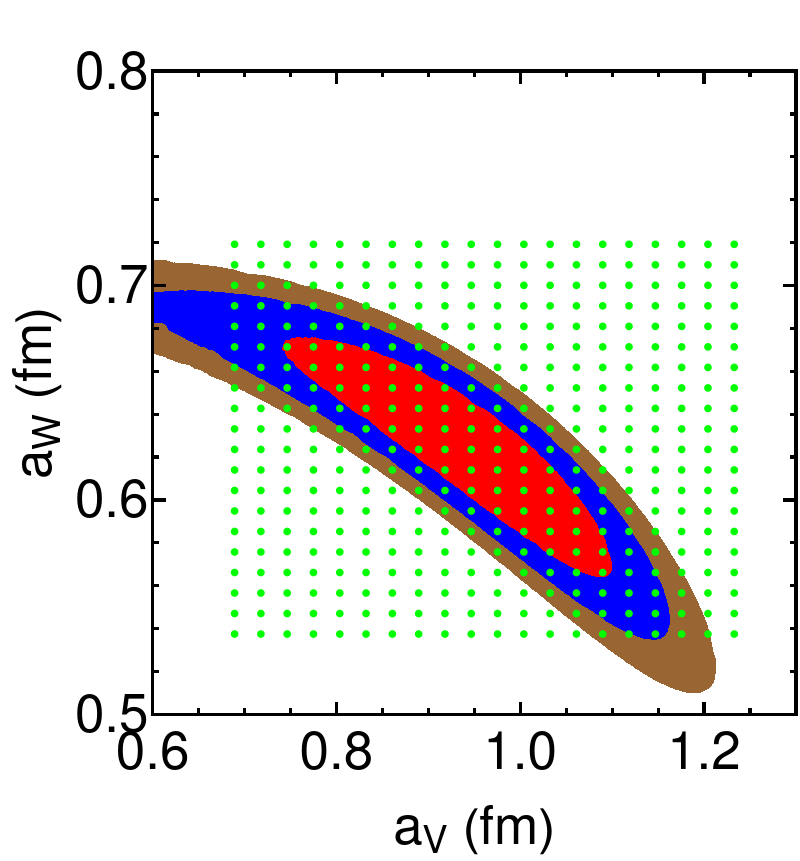}
    \caption{$\chi^2$ contour plots of   $\chi^2$ showing 1$\sigma$, 2$\sigma$, and 3$\sigma$ statistical significance regions for the calculations with three sets of data and $r_0 = 1.1~\text{fm}$. The sampling points are represented by black points for $-3 \leq n_j \leq 3$, and by green points for $-10 \leq n_j < -3$ and $3 < n_j \leq 10$.}
    \label{fig:chi2SP_r11}
\end{figure}

For the sampling points, we will consider as central values of the parameters those
determined by the $\chi^2$ fit, given by the vector $\hat a$, with M components $\hat a_j$. Note that $M=4$ in our case, and  $\hat a_j$ are the best fit values of  $V, a_V, W, a_w  $ given in Table $\ref{tab:Hesse_fix_r0}$. Then, we consider deviation steps of these parameters,
given by the vector $\delta$ with components $\delta_j$, so that when individually added to $\hat a$ they produce a significant deviation in $\chi^2$, given by $\chi^2 \simeq \chi^2_m + 1$. This is achieved taking $\delta_j = \sqrt{2/H_{jj}}$, where $H_{jj}$ are the diagonal elements of the Hessian matrix. The sampling points are labeled by a discrete index $\ell$, and characterized by parameters given by the vector $a^\ell$, so that $a^\ell_j = \hat a_j + n^\ell_j \delta_j $. The set of M integer numbers $n^\ell_j$ determine the sampling point $\ell$. We have chosen  $-10 \le n^\ell_j \le 10$, which produces a total of $21^M$ sampling points. The surfaces are presented along with the sampling points in figures \ref{fig:chi2SP_r11}. We can see that these sampling points cover sufficiently the region contained within the $\chi^2[1 \sigma]$ and  $\chi^2[1 \sigma]$ curves, with the only exception of the very elongated $V-a_V$ contour plot.   Figures \ref{fig:chi2SP_r11} represent the $\chi^2$ surface for the three sets fit, with $r_0=1.1$ fm. We have also evaluated the $\chi^2$ surface for the three sets fit with $r_0=1.2$ fm, and the results are very similar, except for a re-scaling of $V, W$ consistent with Igo ambiguity \cite{igo}. The four sets fit produces qualitatively similar $\chi^2$ surfaces.

\begin{figure}[ht]
    \centering
    \includegraphics[width=0.33\linewidth]{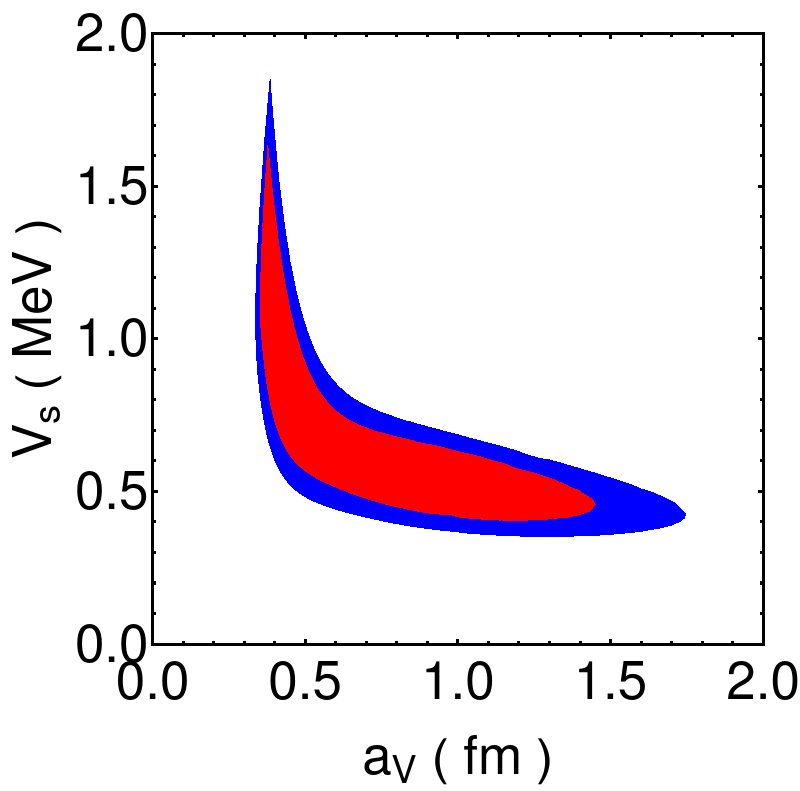}
    \includegraphics[width=0.33\linewidth]{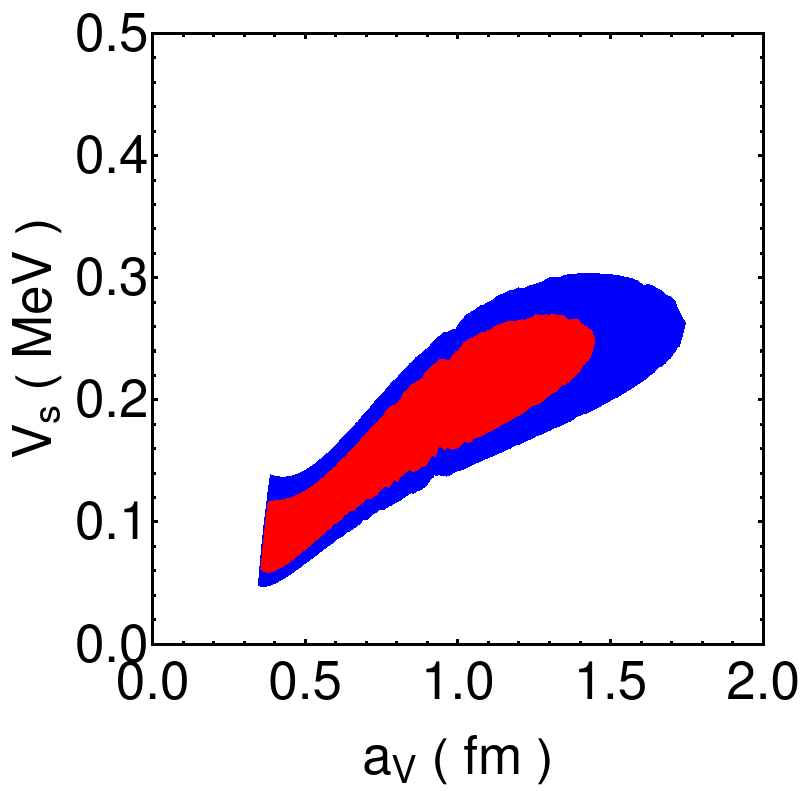}\\
    \includegraphics[width=0.33\linewidth]{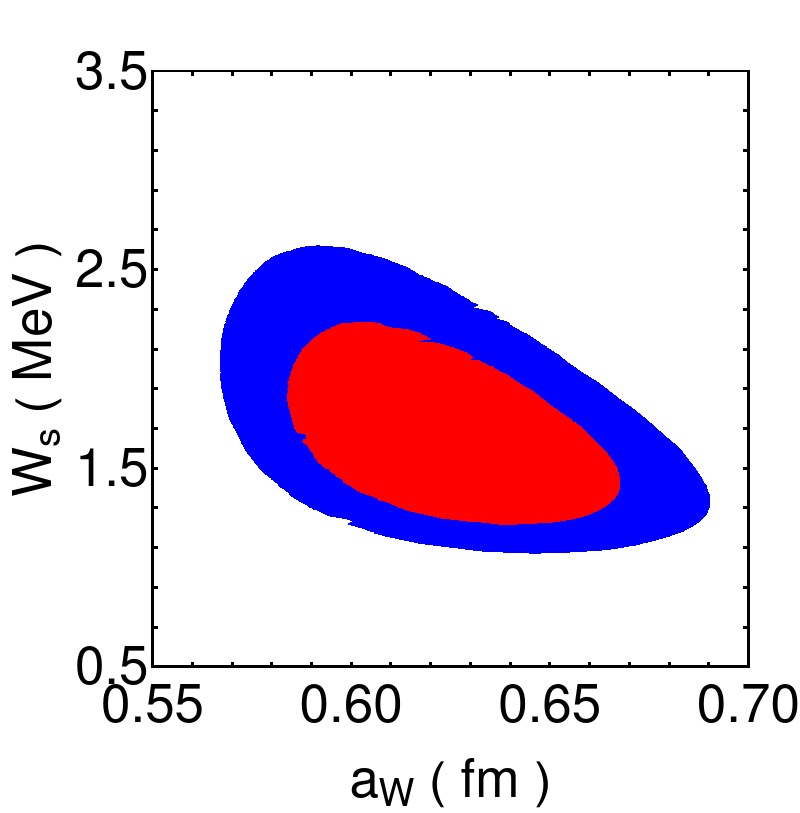}
    \includegraphics[width=0.33\linewidth]{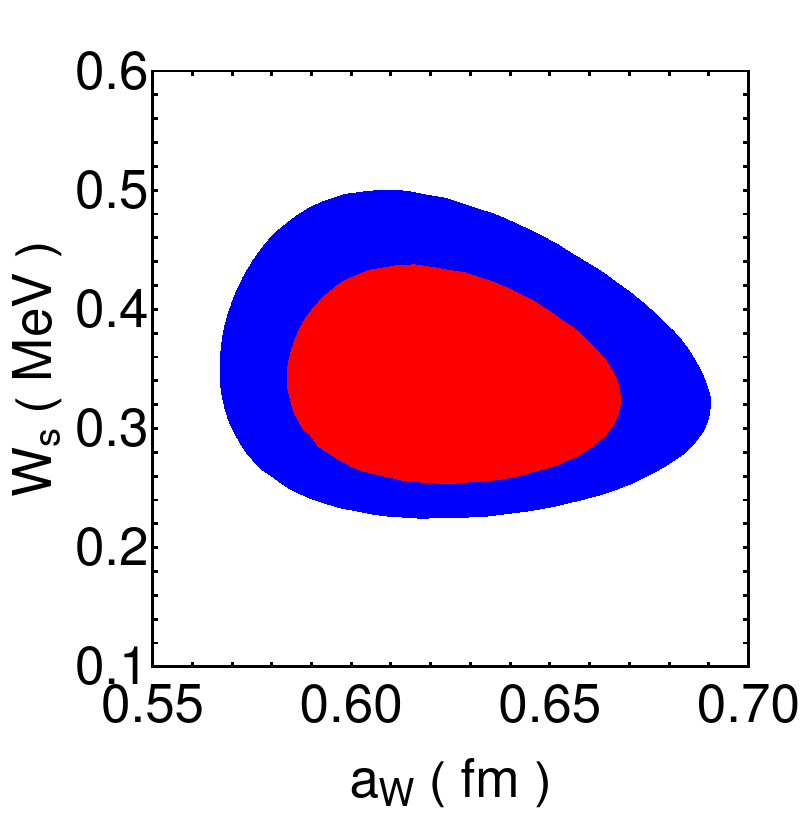}
    \caption{Contour plot of $\chi^2$ showing 1$\sigma$, 2$\sigma$, and 3$\sigma$ statistical significance regions in terms of (Vs, ar) (top) and (Ws, a$_i$) (bottom). On the left side R$_s$ = 9~fm and on the right side R$_s$=10~fm. $r_0=1.1$ fm.}
    \label{fig:Var_r11}
\end{figure}

The strong correlation found between the potential depths $V,W$, and the diffuseness parameters $a_V, a_W$ can be understood plotting the $\chi^2$ surfaces in terms of the potentials $V(r), W(r)$ evaluated at distances that are considered particularly significant for the elastic scattering. 
In figures \ref{fig:Var_r11} we considered   $r=9$ and $r=10$ fm, which could roughly be in the region of the strong absorption radius.  The plots represent the $\chi^2$ contour $V(r)-a_V$ and $W(r)-a_W$. Notice that the contour $V(r)-a_V$ can be understood as a geometrical distortion of the plots $V-a_V$, because  $V(r)$ is a function of $V$ and $a_V$. 
The elongated contour plots $V-a_V$, indicating large correlations, become more straight when plotted as  $V(r)-a_V$, indicating that the correlation is greatly reduced. The same is relevant for the $W(r)-a_W$ plot compared with the $W-a_W$ plot. 
Figures  \ref{fig:Var_r11} indicate that the imaginary potential has, within $1 \sigma$ significance,  a well determined range of values 1.5-2.0 MeV at 9 fm, and 0.3-0.4 MeV at 10 MeV. However, the real potential is significantly smaller and more uncertain: 0.5-1.5 MeV at 9 fm, and 0.1-0.3 MeV at 10 fm. 

\subsection{Optical potential uncertainty bands}

\begin{figure}[H]
    \centering
    \includegraphics[width=0.49\linewidth]{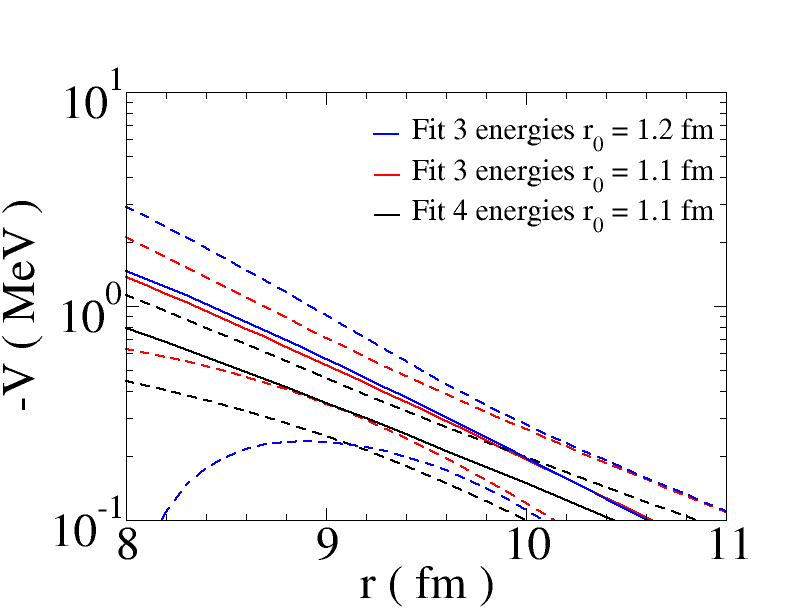}
    \includegraphics[width=0.49\linewidth]{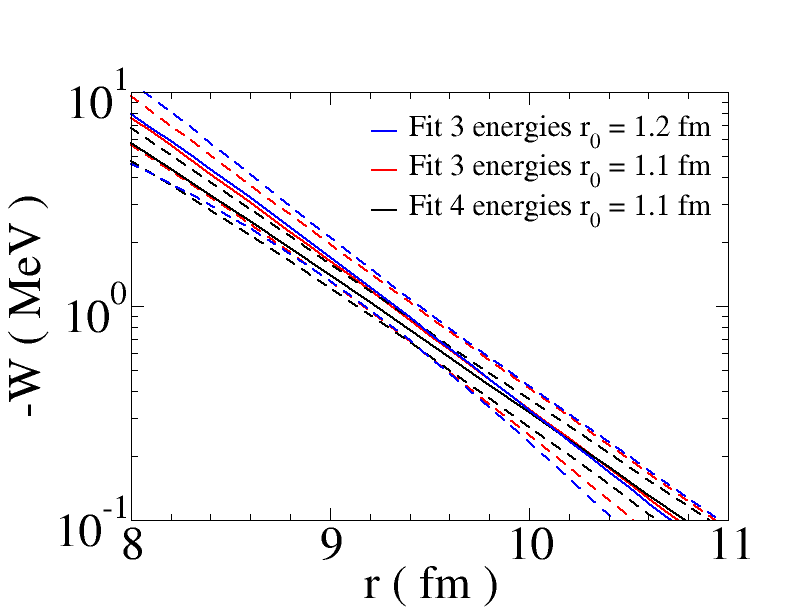}
    \caption{Real and imaginary optical potentials. Full lines represent the calculations for three energies with r$_0$=1.1 and 1.2~fm and for four energies with  r$_0$=1.1~fm
    Dashed lines represent the uncertainty  bands of 1$\sigma$  significance.}
    \label{fig:errorPlotr1112}
\end{figure}

The effect of the correlation become clear when one evaluates the uncertainties of $V(r)$ and $W(r)$. The uncertainty band, for $1\sigma$ and $2\sigma$ significance, obtained from eqs. \ref{eq.sigmabO} and  \ref{eq.sigmaLSO.deriv} , using the analytic derivatives of the Woods-Saxon potentials with respect to the parameters.   In figure \ref{fig:errorPlotr1112} we show, in log scale, the values of  $V(r)$ and $W(r)$. We restricted our plot to the radial range 8-11 fm, because the potentials in the radial range beyond it are irrelevant for the elastic scattering. We see that the range 9-10 fm is where  $V(r)$ and $W(r)$ are better determined, while beyond this range, the uncertainties strongly increase. We can see here the dominance of the imaginary potential over the real potential. Also we see that the uncertainties of the real potential are much larger than those of the imaginary potential. 

\begin{figure}[H]
    \centering
    \includegraphics[width=0.49\linewidth]{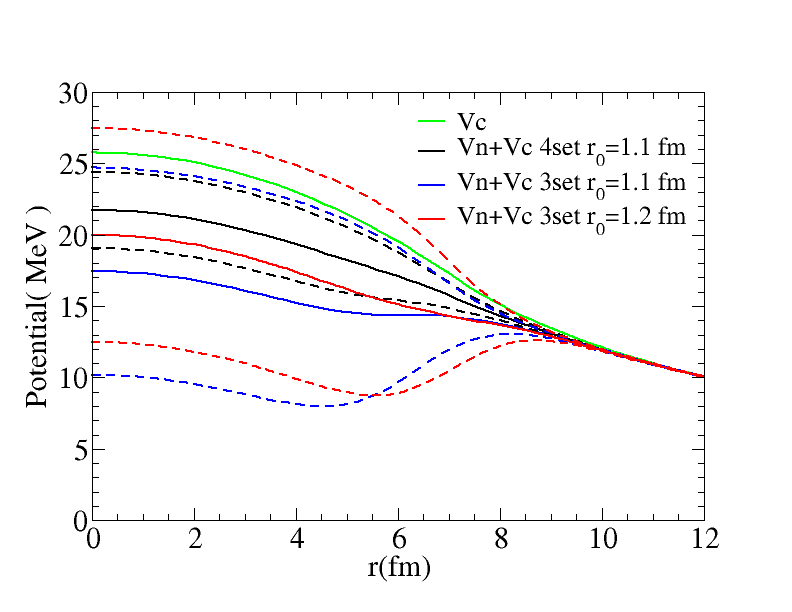}
    \caption{Coulomb plus nuclear potential. Calculations are shown in different colors for three energies with r$_0$=1.1 and 1.2~fm and for four energies with  r$_0$=1.1~fm. The one-sigma uncertainty bands are presented by dashed lines of the same color. The pure Coulomb potential is also shown}
    \label{fig:VnVn_r11}
\end{figure}

To explore further the uncertainty of the real optical  potential, we plot in figure \ref{fig:VnVn_r11} the optical potential added to the Coulomb potential, over a wider radial range, in linear scale. Note that for larger radii, $r > 10$ fm, the real optical potential is completely drowned by the Coulomb potential. For smaller radii, the uncertainties in the real potential are very large. The attractive nuclear potential is so uncertain that one cannot state that it is sufficiently attractive to overcome the repulsive Coulomb potential, generating a Coulomb barrier.  Note that the calculation with four energies produces a significantly weaker nuclear potential, that is not able, within the $1\sigma$ band, to  compensate the Coulomb repulsion, generating a barrier. 

These results are sobering. Elastic differential cross section,   from the present set of scattering cross section data, are not sufficient to determine, accurately and independently, the real and the imaginary optical potentials. Notice that the main feature that is displayed by the elastic differential cross sections at energies around the Coulomb  barrier is a reduction with respect to Rutherford, and that can be achieved both by absorption (imaginary potential) and by diffraction (real potential). Indeed, other more accurate sets of elastic differential cross sections may be able to determine independently real and imaginary potentials. However, for that purpose, they should be able to display accurately features that depend differently on the real and the imaginary potentials. This is the case of the rainbow, this is, the increase of the differential cross section over the Rutherford value, that happens at the rainbow angle. That feature, measured accurately in scattering of stable nuclei such as $^4$He and $^{16}$O, has larger uncertainties in unstable nuclei such as $^8$Li. 
Alternatively, one may assume that the real optical potential can be obtained microscopically, for example by double folding. The expectation is that any reasonable real potential, when complemented with a properly adjusted imaginary potential, should be able to reproduce the elastic scattering data. However, the agreement should not be taken as any evidence of the goodness of the microscopic real potential.
Our results highlight the difficulty of properly investigating sharp energy dependences of the optical potentials around the Coulomb barrier, which are needed to explore dispersion relations.

\subsection{Differential cross sections uncertainty bands}

\begin{figure}
    \centering
        \includegraphics[width=0.40\linewidth]{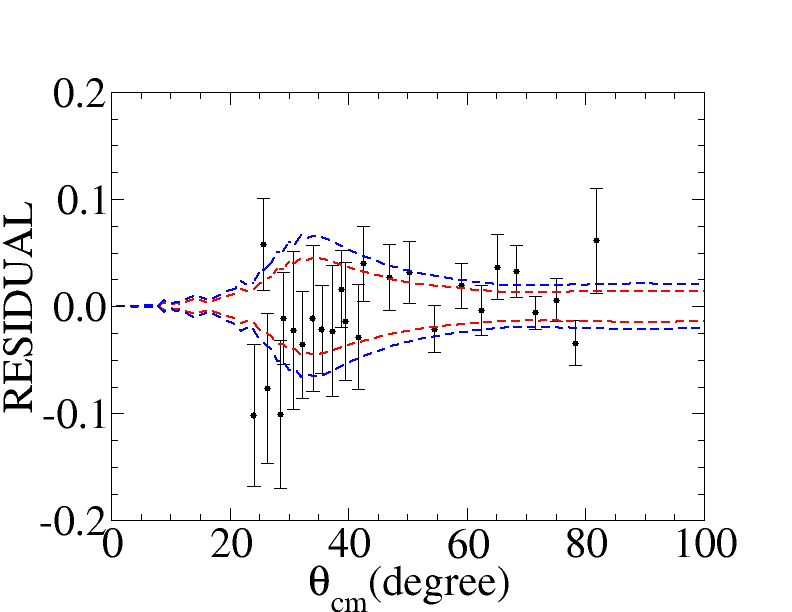} 
        \includegraphics[width=0.40\linewidth]{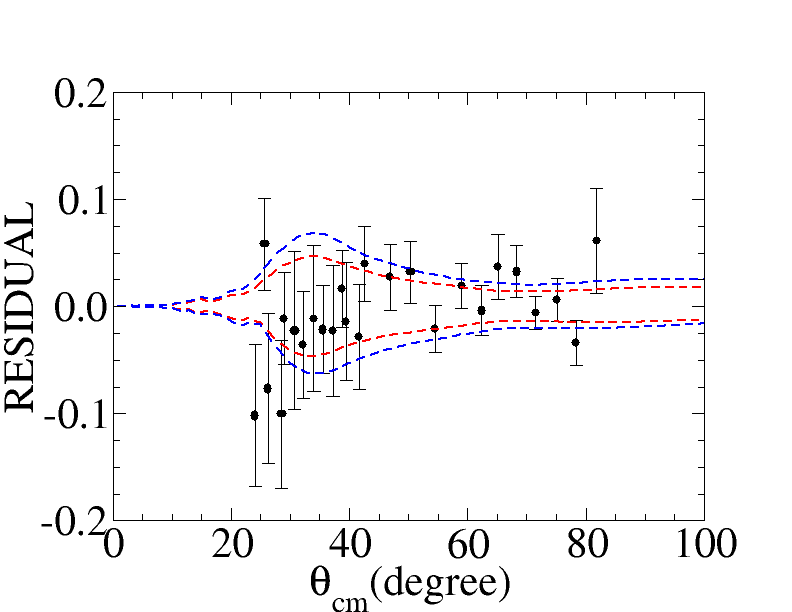} \\
        \includegraphics[width=0.40\linewidth]{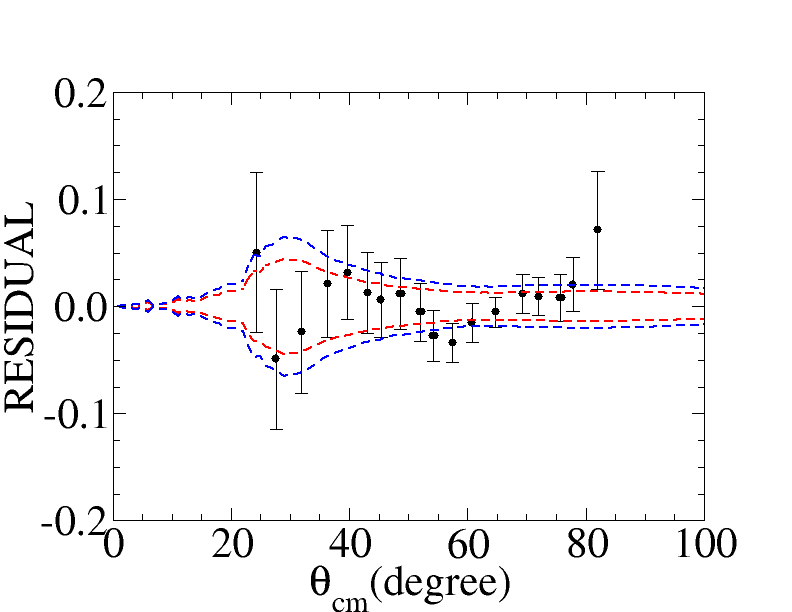}
        \includegraphics[width=0.40\linewidth]{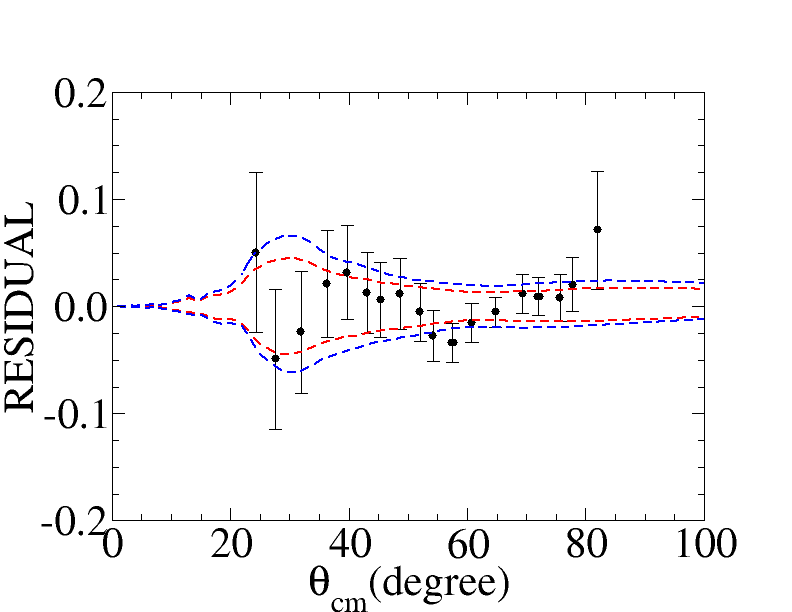}\\
        \includegraphics[width=0.40\linewidth]{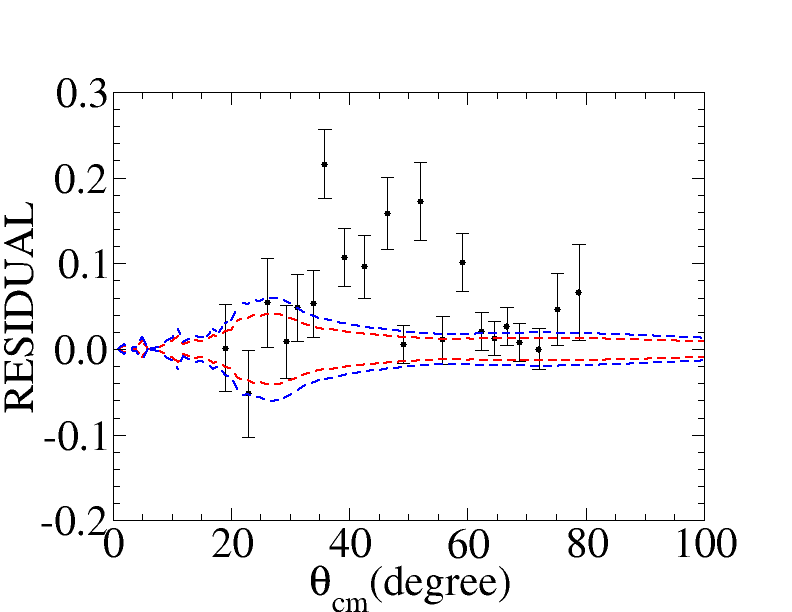}
        \includegraphics[width=0.40\linewidth]{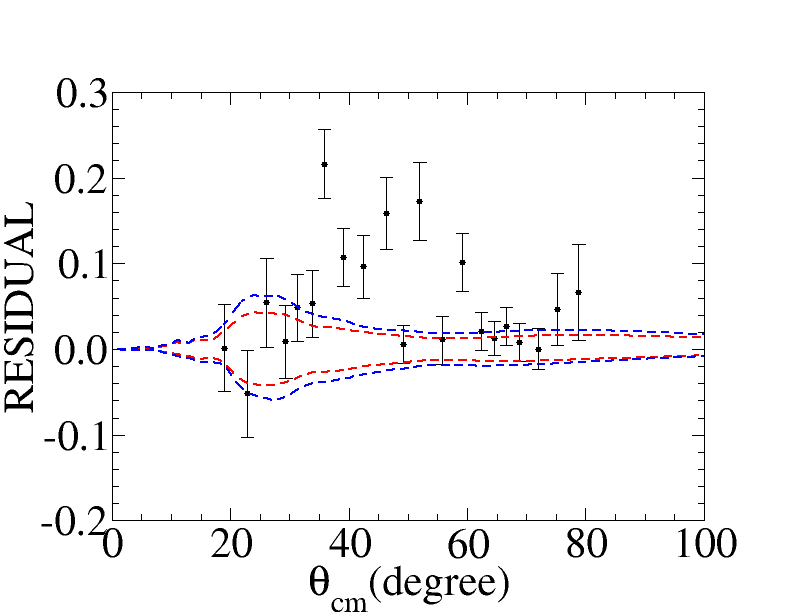}\\
        \includegraphics[width=0.40\linewidth]{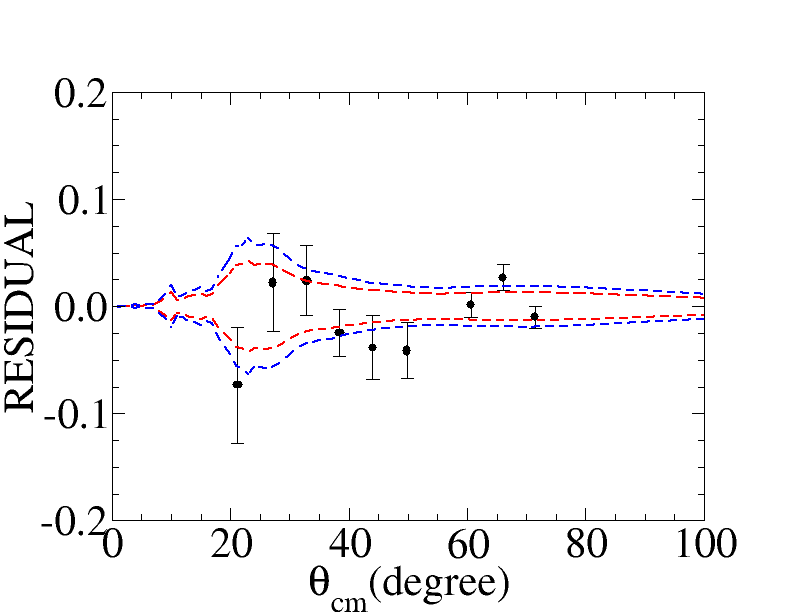} 
        \includegraphics[width=0.40\linewidth]{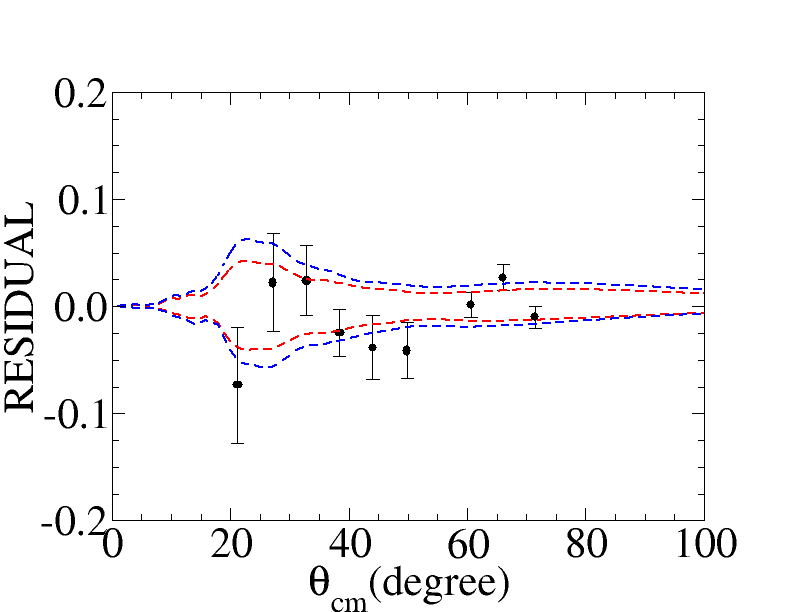}
    \caption{Residuals for 23.9, 26.1, 28.7  and 30.0 MeV (from top to bottom), compared with the  1$\sigma$ and 2$\sigma$ uncertainty bands,  for the three set fit with $r_0 = 1.1~\text{fm}$. Left panels  use the simplified approach and the right panels use the general approach.
    }
    \label{fig:residuals_r11}
\end{figure}

We present here the uncertainty bands of the differential cross section calculations, considering the three sets of experimental data at the energies  23.9, 26.1  and 30.0 MeV. We consider the bands corresponding to $1\sigma$ and $2\sigma$ statistical significance.
We use the two approaches described in this work. In the general approach, we consider all the sampling points which are within the  $1\sigma$ and $2\sigma$ regions, and obtain the maximum and the minimum value of the differential cross sections at each scattering angle and energy. The general approach is very robust, as it does not require any assumption about the dependence of the cross sections on the parameters. it is conceptually very simple, but requires to perform hundreds of calculations. This is not a problem in this case, as the calculations required are optical model calculations, although it may be problematic when applied to more complex scattering calculations.
The simplified approach makes use of the covariance matrix of the parameters, which is obtained from the fitting codes, and applies the uncertainty propagation formulae given in eqs. \ref{eq.sigmaLSO.delta} and \ref{eq.sigmabO}. The simplified method assumes smooth dependence on the cross sections on the parameters, but requires very few calculations: only 2M, this is eight calculations, to determine the variation of the cross sections with the parameters. The results of both methods on the elastic differential cross sections are shown in figures \ref{fig:residuals_r11}, for the four energies for which there are experimental data,  23.9, 26.1, 28,7  and 30.0 MeV.

We have extracted the residuals, this is, the difference between the experimental points of the differential cross sections divided by Rutherford,  and the best fit calculations. We compare these residuals with the $1\sigma$ and $2\sigma$ uncertainty bands in figure \ref{fig:residuals_r11}.
The uncertainty bands have been obtained using the simplified approach (left), and the general approach (right). The results are very similar, although one can see some differences. In particular, while in the simplified approach the bands are symmetric, in the general approach the bands, determined by the maximum and the minimum calculations of the sampling points,  are not symmetric in general.
. 
We see that the residuals of the experimental points at 23.9, 26.1  and 30.0 MeV  are consistent with the theoretical uncertainty bands. More specifically, taking the points at all energies,  from 52 experimental points, 26 experimental points (50\%) are within the $1\sigma$ uncertainty bands, while 37 experimental points (71\%) are within the $2\sigma$ uncertainty bands.  A naive statistical perspective could lead to the expectation that  $1\sigma$ band should contain 62\% of the points and the    $2\sigma$ band should contain 95\% of the points. However, this relation is not strict. We should remember that the bands of the calculations are associated to conditional probability that calculations take some values ${\bf given}$ the experimental data, which are different from the probability that the experimental points take some value  ${\bf given}$ the results of the calculations. This difference in conditional probabilities is the essence of the Bayesian treatment. Nevertheless, it is reassuring that the calculated uncertainty bands are reasonably consistent with the data.


The experimental points at 28.7 MeV are not consistent with the calculations, even when theoretical uncertainty bands are considered.  There, 11 data out of 21 are outside the $2\sigma$ uncertainty band,  This indicates that these experimental data cannot be described by the energy independent optical potential which describes the other energies.

\subsection{Robustness of the method}

We have seen, from figure \ref{fig:el}, that there are  three sets the experimental data, at the energies   23.9, 26.1  and 30.0 MeV, that are well fitted by optical model calculations. Let us candidly call them `nice' data. However,  experimental at the energy  28.7 MeV, that cannot be fitted with the same quality. We can call them `nasty' data. The application of the method  with three sets  uses only the `nice' data, producing very good fits characterized by $\chi^2_m$ values similar to the number of degrees of freedom. The application of the method with with four sets uses both `nice' and `nasty' data, and generates worse fits, with $\chi^2_m$ values larger than the number of degrees of freedom.
 Here we want to evaluate what is the effect in the evaluation of the uncertainty bands of introducing `nasty' data, which present significant deviations from the model fits. This is important, because in an actual application of this method to various sets of real data, we may have `nice' as well as `nasty' data, with different quality of agreement with the model calculations. 
`Nasty' data could appear due to a variety of reasons, from the adequacy of the model, to some unforeseen experimental issue. We want to check if the method is robust with respect of the experimental data included, so that the results to not strongly depend on a prior separation of `nice' from `nasty' data.   This separation could be difficult, arbitrary, and it could hide unexpected physics.
For that purpose, we compare the results using the three sets of `nice' data (this is, 23.9, 26.1  and 30.0 MeV), with the results using four sets, both `nice' and `nasty' data. 

In figure \ref{fig:el} we present the best fits for three sets of data and four sets of data. We see that the results of the differential cross section calculations are very similar between the three sets and the four sets, with differences that are well within the experimental error bars. In figure \ref{fig:errorPlotr1112} we see that the optical potentials $V(r), W(r)$ obtained with the three sets of data and the four sets of data are rather similar, with differences in the relevant radial region that are well within the deduced $1\sigma$ uncertainties. In figure \ref{fig:VnVn_r11} we see that the real potential obtained from the fit to the four sets of data is significantly weaker than that one from the three sets of data, so that it is unable to overcome the Coulomb repulsion. Thus, the inclusion of the `nasty' data makes the fitted real potential less physical.

\begin{figure}
    \centering
        \includegraphics[width=0.40\linewidth]{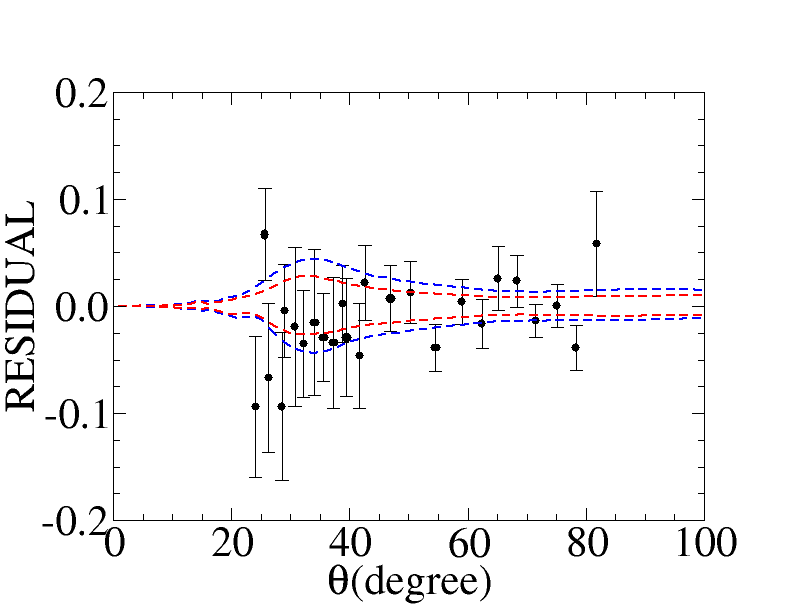}
        \includegraphics[width=0.40\linewidth]{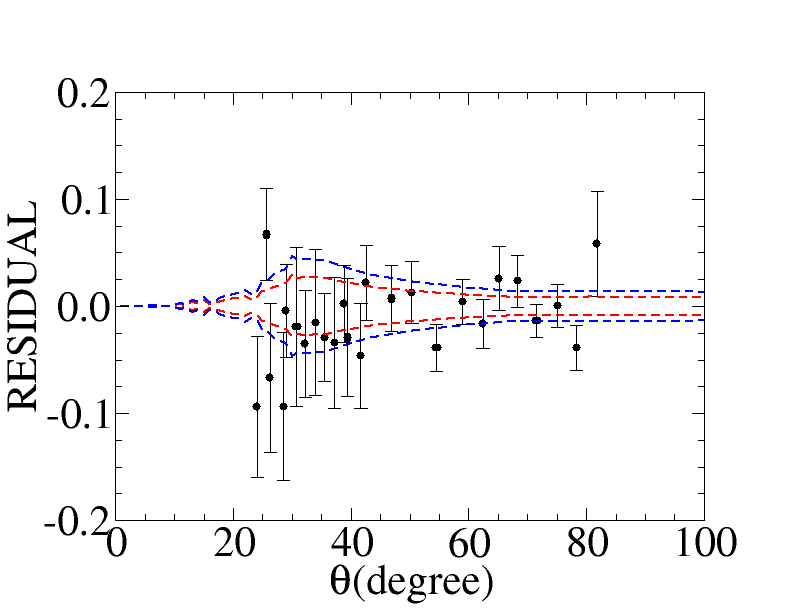}\\
        \includegraphics[width=0.40\linewidth]{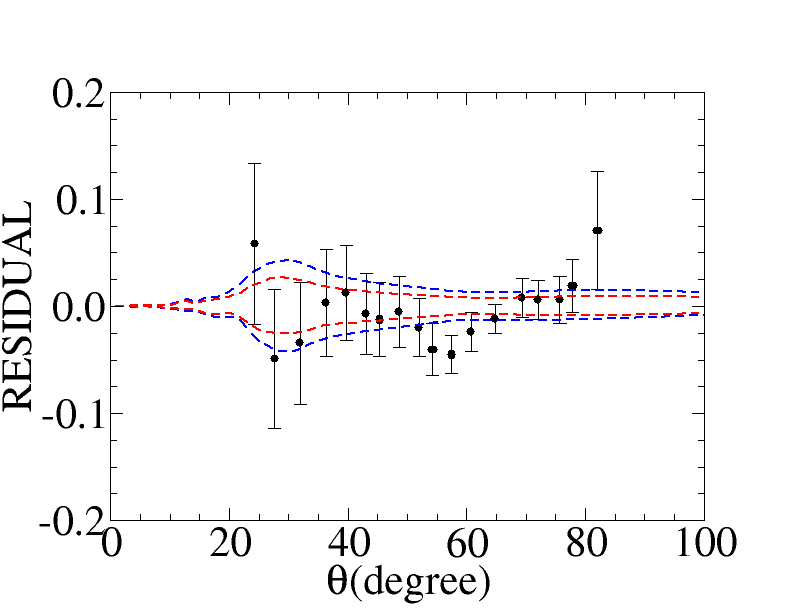}
        \includegraphics[width=0.40\linewidth]{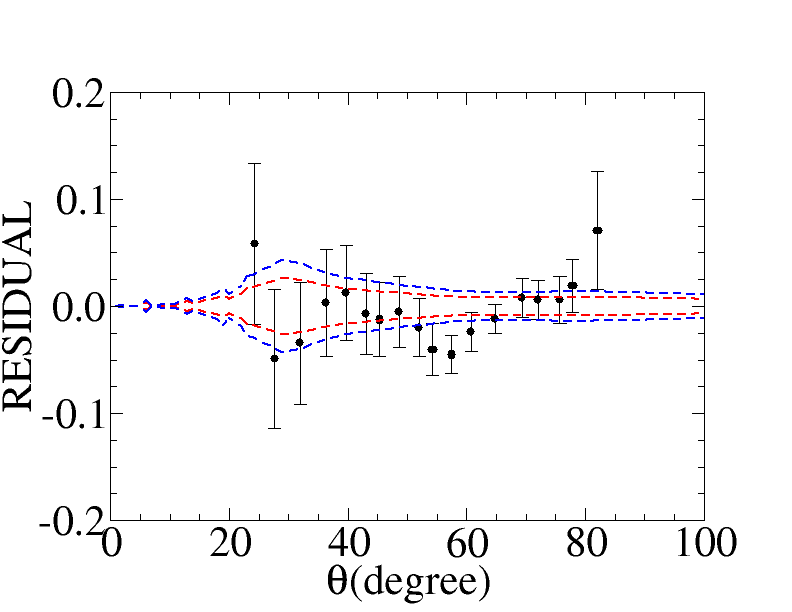}\\
        \includegraphics[width=0.40\linewidth]{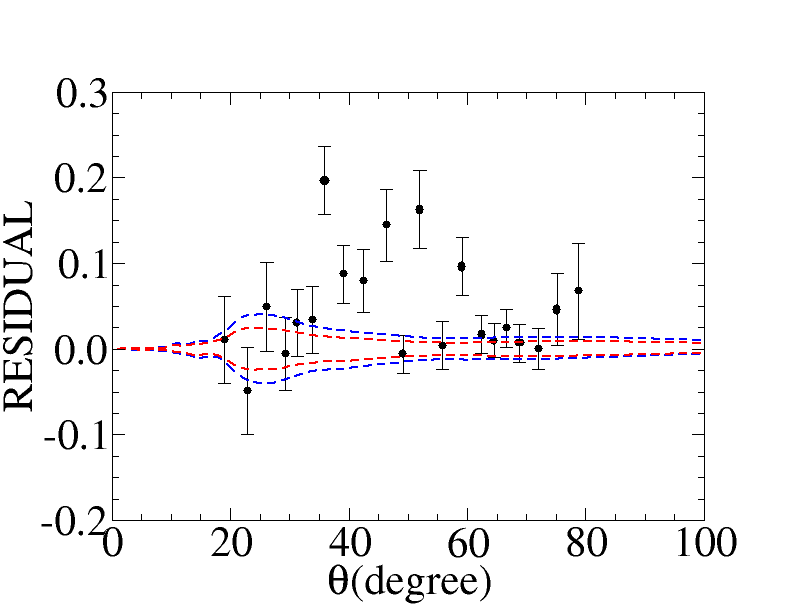}
        \includegraphics[width=0.40\linewidth]{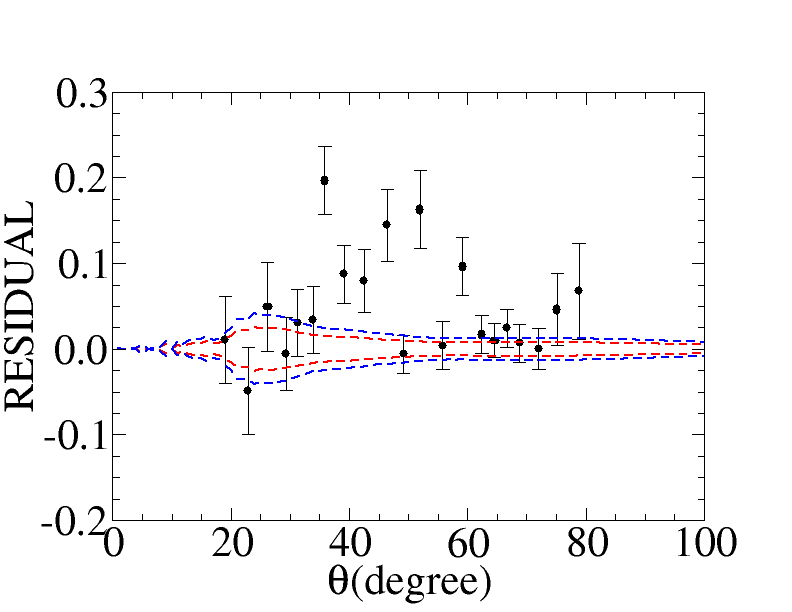}\\
        \includegraphics[width=0.40\linewidth]{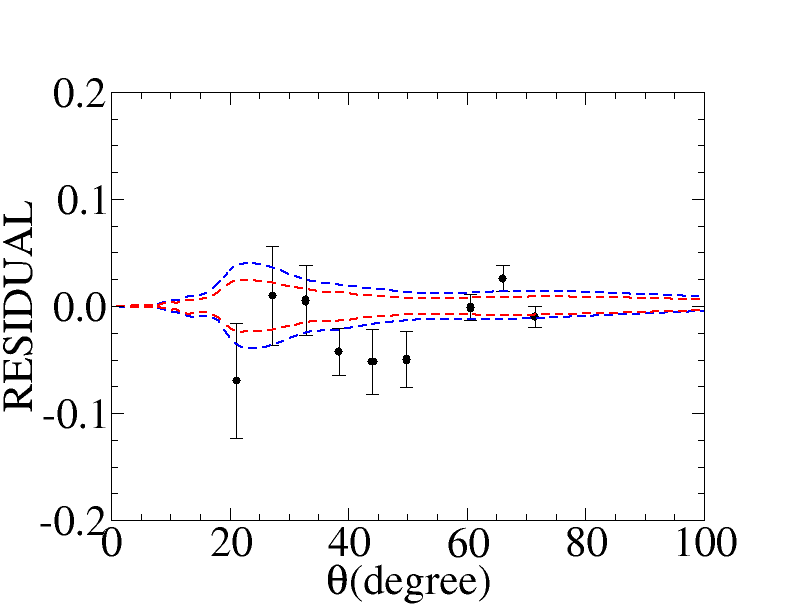}
        \includegraphics[width=0.40\linewidth]{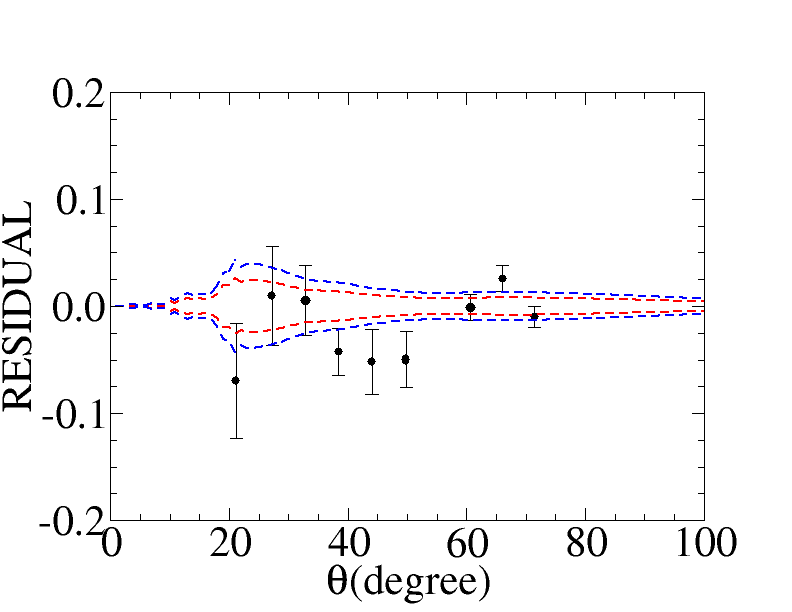}\\
    \caption{Residuals for the optical model fit using the 4 sets of data, compared with the $1\sigma$ and $2\sigma$ bands, obtained in the simplified (left) and in the general (right) approaches.}
    \label{fig:diff287}
\end{figure}

We want to see the effect of the uncertainty bands of the calculated cross sections. In figure \ref{fig:residuals_r11} the residuals of the experimental points, and the uncertainty bands of the calculations, are presented for the three sets, this is, including only `nice data' in the method. There one can evaluate uncertainty bands for different energies, and one sees that there are strong deviations of the experimental data only for the `nasty' data at 28.7 MeV. This is not surprising, as these data were not included in the method. However, in figure \ref{fig:diff287} we present the results using the four sets of data, which includes both the `nice' and the `nasty' data. There we can see that the uncertainty bands are somewhat more narrow, compared to figure \ref{fig:residuals_r11}. This is partly due to small differences in the covariance matrix of the parameters, and also to the fact that the enhancement factors in table \ref{tab:sigUnce_fix_r0} are smaller when the `nasty' data are included.

The results of figure \ref{fig:diff287} indicate that the inclusion of the `nasty' data affects slightly the cross section calculations and more significantly the uncertainty bands. However, both calculations (with and without `nasty' data), contain  reasonably the three sets of  `nice' data within uncertainty bands, while the  `nasty' data deviates from the band.  Thus, we can reach the robust conclusion that the data at 28.7 MeV is not consistent with optical model calculations based on an energy independent potential, considering explicitly the uncertainties.  A proper uncertainty analysis should disregard the 28.7 MeV data for the fit, so that the proper uncertainty bands are those of the calculations with the three sets of data.

It seems counter-intuitive that calculations with  `nasty' data, and hence worse fits (larger values of $\chi_m^2/L$) produce smaller uncertainties in the calculations. This is a direct consequence of Bayes treatment. The probability distribution of the parameters is given by  Eq. \ref{Bayesp}, where the probability $P(\chi^2_m,L)$ appears in the denominator. Thus, when $\chi^2_m > L$, $P(\chi^2_m,L)$ could be a very small number. This means, in Bayesian terms, that the probability $P(H)$ of obtaining the data, is small. However, the probability of the parameters, given the data, is given by the ratio $P(b, L)/P(\chi^2_m,L)$, which gets narrower as a function of $b-\chi^2_m$, when $\chi^2_m$ increases, leading to more narrow error bands for the parameters, and hence for the calculations. This indicates that caution is required when Bayesian methods, such as this one, are used. The model, along with the data, should be assessed, according to the value of $P(H)= P(\chi^2_m,L)$. Once the model is accepted as a reasonable description of the data, the uncertainty of the parameters are determined by the ratio $P(b, L)/P(\chi^2_m,L)$, In our case, this assessment led to the consideration that the model (optical model calculations with four parameters) is reasonable for three sets of data, not four. This leads to the consideration of the uncertainty bands given in figure \ref{fig:residuals_r11}, not those in  figure \ref{fig:diff287}.

\section{Summary, discussion and conclusions}
\label{sec:summary}

We present a method to describe the uncertainties in the calculations of scattering magnitudes for nuclear reactions, associated to the optical model parameters. The procedure starts with standard $\chi^2$ fits, which determine the least square values of the optical model parameters $\hat a$ as well as the covariance matrix. Then, in the spirit of the maximum likelihood method, the parameter space is structured in regions characterized by the $\chi^2(a)$ surface. The statistical significance and p-values  of these regions are determined using Bayes theorem. The uncertainty bands of arbitrary magnitudes derived from the parameters, such as optical potentials and scattering magnitudes are evaluated using two approaches: In the general approach, the uncertainty band is obtained considering the maximum and minimum value of the magnitude in the sampling points belonging to the relevant parameter region. In the simplified approach, the uncertainty band is determined by propagation of the magnitudes in terms of the parameters, making use of the covariance matrix of the parameters, and a factor $f[b]$ associated to the statistical significance.

We have applied this procedure to investigate the uncertainties in the optical potentials and in the differential cross section of the system $^8$Li + $^{58}$Ni. We consider separately three data sets at energies~\cite{data}, 23.9, 26.1 and 30.0 MeV, for which good optical model fits are obtained, and four data sets,  including also the data at the energy 28.7 MeV, for which the optical model fit is not as good. 
We found a satisfactory fit of the data of the three data sets, using energy-independent parameters. The four data sets have can also be fitted, but the quality of the fit is worse.

We have evaluated numerically the $\chi^2(a)$ surfaces, and represented its contour plots for each pair of  parameters. We evaluate $1\sigma$, $2\sigma$ and $3\sigma$, contour plots,  that display  concentric  shapes, only roughly ellipsoidal close to the minimum. 
Our results contrast with the rather irregular contour plots presented in figure 2 of reference \cite{Nunes2019} for the Bayesian calculation. The reason for this discrepancy is not clear to us. 
A benchmark calculation comparing different methods to the same set of data would be in order.

We have evaluated the  $1\sigma$ and  $2\sigma$ uncertainty bands of the optical potentials $V(r), W(r)$ as a function of the distance, making use of the covariance matrix of the parameters and error propagation. We find that these potentials are reasonably well determined in the range $r=(9, 10)$ fm. Beyond that range, specially the real potential, is very uncertain. This indicates that the adequacy of microscopic real potentials should not be taken as a proof of the quality of the interaction, but rather as a consequence of the weak dependence of the scattering observables with the real potential. The uncertainty bands of the imaginary potentials are more narrow, indicating that elastic scattering observables are indeed sensitive to the imaginary potential.

We have evaluated the  $1\sigma$ and  $2\sigma$ uncertainty bands of the differential cross sections for the energies
23.9, 26.1, 28.7 and  30.0 MeV, as a function of the scattering angle, using the general approach and the simplified approach.
Both calculations give very similar results. 
The experimental data at 23.9, 26.1, and 30.0 MeV are consistent with the uncertainty bands of the optical model calculations. However, the data at 28.7 MeV are not consistent with the optical model calculation, even when the uncertainty of the calculation is included. We have evaluated the robustness of the method, considering the effect of including the 28.7 MeV data which are poorly fitted by the optical model calculations. We find that, although the global $\chi^2$ values are strongly increased,  the calculated values are very similar, and the uncertainty bands of the calculations are reduced. This indicates that the method can be applied systematically, without a previous filtering of the experimental data.

The overall consistency with the simplified approach and the general approach allows to understand the relation between {\em frequentist} and Bayesian uncertainties, shown in \cite{ Nunes2019}.  Our calculations indicate that the main difference is due to the enhancement factors $f[p]$. The {\em frequentist} approach could be associated to the least square method, mostly used in search routines such as SFRESCO \cite{SFRESCO}. However, the uncertainties that come from these codes are a standard deviation, and they can only be assigned an statistical meaning when multiplied by the factors $f[p]$.  Conversely, the general approach that we have discussed here corresponds to a true Bayesian treatment, where we have chosen equally spaced sampling points instead of a Monte-Carlo sampling method. Thus, we can conclude that the main difference between {\em frequentist} and Bayesian  approaches, is the need to introduce in the former the enhancement factors  $f[p]$. Indeed, the factors $f[p]$ have a Bayesian origin, as they were derived using Bayes formula.

The uncertainty bands in the elastic differential cross sections can  be useful for planning experiments. The bands, that reflect the uncertainties in the optical potential parameters, are wider in some angular ranges, especially around the rainbow angle, and more narrow at other angles. New experimental measurements at a given angle would be effective to narrow the uncertainty bands of their expected experimental uncertainties are smaller than the uncertainty band at that angle. Thus, experiments covering the rainbow angle can be more effective to determine better the optical potential, provided that their uncertainties are small enough. A design of the angular coverage of the detectors for a new experiment, which provides new elastic data to improve the knowledge of the optical potential, can benefit from the previous knowledge of the uncertainty band of the elastic cross section, deduced from previous data.

We expect that this method, in the general or simplified approach, could be used to evaluate uncertainties in complex,  state of the art reaction calculations. Let us illustrate the application of the simplified method for a complex scattering calculation, such as a CDCC calculation for $^{11}$Be + $^{197}$Au, producing elastic, inelastic and break-up calculations  \cite{11be197au}. $^{11}$Be is described as a core of $^{10}$Be coupled to a neutron. The CDCC calculation involves n+ $^{10}$Be and n +  $^{197}$Au potentials, which are usually well determined through the structure of $^{11}$Be,
and from systematics on neutron optical potentials. However, the core-core interaction  $^{10}$Be + $^{197}$Au has to be determined by  $^{10}$Be + $^{197}$Au  scattering data. 
We can envisage the determination of 4 parameters ($V, a_V, W, a_w$) in the core-core interactions, from the best $\chi^2_m$ fit to elastic  $^{10}$Be + $^{197}$Au data.  We can perform a CDCC calculation with these parameters, and that gives us the  maximum likelihood value for the elastic, inelastic and break-up differential cross sections for $^{11}$Be + $^{197}$Au. We can evaluate also the uncertainty in these magnitudes. For that, we consider the covariance matrix, which gives the standard deviation of the parameters ($\sigma(V), \sigma(a_V), \sigma(W), \sigma(a_w)$).
Then one perform 8 extra CDCC calculations, changing the parameters to ($V \pm \sigma(V), a_V \pm \sigma(a_V), W\pm \sigma( W), a_w \pm \sigma(a_W)$), obtain the changes of the elastic, inelastic and break-up  differential cross sections, with respect to the parameters, and perform error propagation following eq. \ref{eq.sigmabO}. 

\begin{acknowledgments}
The authors acknowledge the contribution  of Antonio Moro for the advice with the use of SFRESCO and Mario Gomez for a critical reading of the manuscript. 
We acknowledge support by Horizon Europe Grant Agreement No 101057511 (EURO-LABS - EUROpean Laboratories for Accelerator Based Sciences), and by spanish AEI project PID2021-123879OB-C21 (Física con neutrones, instrumentación nuclear y desarrollos para hadronterapia en el CNA e instalaciones internacionales)

This work has been partially supported by Funda\c{c}\~ao de Amparo \`a Pesquisa do Estado de S\~ao Paulo, Brazil (FAPESP) - contracts ns. 2019/07767-1 and 2021/12254-3.

\end{acknowledgments}

\appendix
\section{Least square, Maximum likelihood and  Bayesian methods}
\label{ap:leastSquare}

In this appendix we present a derivation of the relations of Least Square, Maximum Likelihood and Bayesian methods which are required for a detailed understanding of the method. We will also derive the expressions required to obtain the probabilities of the different sampling points, based on a Bayesian treatment.  It should be noticed that the main objective of the paper is to obtain uncertainties bands, which correspond to the maximum and minimum  values of the observables evaluated in a given parameter region. This objective is independent on the particular values of the probabilities assigned to the sampling points. However, we leave here the expressions of these probabilities, as they are relevant to more detailed statistical properties of the calculations, such as percentile distributions of the observables. 
\subsection{Least square method}.

Here we will review the least square (LS) method, following \cite{cowan98}. This is the method used in standard  $\chi^2$ minimization codes, such as MINUIT~\cite{MINUIT}, which is used, for example in SFRESCO \cite{SFRESCO}. These codes minimize the quantity $\chi^2(a)$, described previously, and so determine the parameters $\hat  a$ that make  $\chi^2(\hat a)= \chi^2_m$. This minimum defines values for the theoretical predictions $y_i(\hat a)$, as well as the first and second derivatives of the theoretical predictions $y_i(a)$ with respect to the parameters $a_j$, $a_k$, evaluated at $\hat a$:
\begin{eqnarray}
    y'_{ij} &=&  \left. {1 \over \sigma_i} {\partial y_i(a) \over \partial a_j }\right|_{a = \hat a} \\ 
    y''_{ijk} &=&  \left. {1 \over \sigma_i} {\partial^2 y_i(a) \over \partial a_j \partial a_k}\right|_{a = \hat a} 
\end{eqnarray}

The parameter vector $\hat a$ is defined by the condition that $\chi^2(\hat a) $ is the absolute minimum. This implies that all the partial derivatives vanish 
\begin{equation}
\left. {\partial \chi^2(a) \over \partial a_j }\right|_{a = \hat a} = 2 \sum_i  y'_{ij} {y_i(\hat a)-y_i^e \over \sigma_i} = 0   \label{partchi2}
\end{equation}
The Hessian matrix $H_{kj}$ is defined by the second derivatives of the $\chi^2(a)$ function
\begin{equation}
   H_{kj} = \left. {\partial^2 \chi^2(a) \over \partial a_j  \partial a_k  }  \right|_{a = \hat a} 
 =2 \sum_i y'_{ik} y'_{ij} + 2 \sum_i   y''_{ikj} {y_i(\hat a) - \hat y_i \over \sigma_i}
\end{equation}
While the first term in the LHS have a positive-definite contribution to the Hessian matrix, the second term contains residual terms $  (y_i(\hat a)-y_i^e) / \sigma_i$ which can be positive or negative. When these residual terms are multiplied by the first derivative $y'_{ij}$, they cancel, following  equation~(\ref{partchi2}).  Here they multiply the second derivative $y''_{ikj}$, so although they do not strictly cancel, their contribution will be small compared to the first term. So, assuming the randomness of the  residual terms,
\begin{equation}
   H_{kj} \simeq 2 \sum_i y'_{ik} y'_{ij} 
\end{equation}

The expression 
\begin{equation}
 \sum_i  y'_{ij} {y_i(\hat a)-y_i^e \over \sigma_i} = 0   \label{partchi2b}
\end{equation}
can be seen as defining $a_m$ as an implicit function of the measurements $\hat y_i$.
As $\hat y_i$ has uncertainties given by $\sigma_i$, error propagation can be carried out to obtain the uncertainties of the parameters. Indeed, if the experimental observations $\hat y_i$ are modified by some small quantity $\xi_i \sigma_i$, then the parameters $\hat a_j$ are modified by small quantities $\alpha_j$, which fulfill
\begin{equation}
2 \sum_{ki }   y'_{ij}y'_{ik} \alpha_k = \sum_k H_{jk} \alpha_k = 2 \sum_i y'_{ij} \xi_i   
\end{equation}

The deviations of the parameters $\alpha_k  $ can be expressed as a linear combination of the deviations of the experimental data, given in terms of $\xi_i$
\begin{equation}
\alpha_k  =  \sum_j H^{-1}_{kj}  \sum_i  2  y'_{ij} \xi_i  = \sum_k c_{ki} \xi_i  \label{deltaak}
\end{equation}
Note that $\sum_i c_{ki} c_{ji} = 2   H^{-1}_{kj} $.
The parameters $\xi_i$ describing deviations of the experimental data have a probability distribution corresponding to N independent normal distributions:
\begin{equation}
    {\cal P}_D(\xi_i) =  {1 \over (2\pi)^{N/2}} e^{- \sum_i {\xi_i^2 /2}}
\end{equation}

The least square covariance matrix of the parameters $cov_{LS}(a_k, a_j)$ can be expressed in terms of the expectation value of the  product $ \alpha_k  \alpha_j  $, in the probability distribution assumed for the experimental deviations $ \delta y_i$, ${\cal P}_D(\delta y_i)$
\begin{equation}
   cov_{LS}(a_k, a_j) = \int d^N\! \xi_i \; {\cal P}_D(\xi_i)  \alpha_k  \alpha_j .
\end{equation}
The normal probability distribution fulfills:
\begin{equation}
   cov(\xi_i, \xi_{i'})  = \delta(i, i')
\end{equation}
The result obtained is
\begin{eqnarray}
cov_{LS}(a_k, a_j) &=& \sum_{ll'} H^{-1}_{kl} H^{-1}_{jl'} \sum_i {4 \over \sigma_i^2}   y'_{il}   y'_{il'} \\ &\simeq& 2 \sum_{ll'} H^{-1}_{kl} H^{-1}_{jl'} H_{l l'} = 2 H^{-1}_{kj} \label{covhess}
\end{eqnarray}

The diagonal elements of the covariance matrix corresponds to the variance of the different parameters, which is the square of the standard deviation of the parameters. Thus, 
\begin{equation}
    \sigma(a_j)^2  = cov_{LS}(a_j, a_j)
\end{equation}
The correlation matrix is obtained from the covariance matrix, divided by the product of the standard deviations 
\begin{equation}
    corr_{LS}(a_j, a_k) = {cov_{LS}(a_j, a_k) \over \sigma(a_j)\sigma(a_k)}
\end{equation}

It should be stressed that the expression of the covariance \ref{covhess} does not take into account any probability information related to the quality of the fit, as described by the $\chi^2$ value. Thus the standard deviation of the parameters $\sigma(a_j)$ obtained from the usual codes, such as SFRESCO, do not contain, a priori, a statistical meaning similar to that of the standard deviation of a normal distribution. Notice that, in deriving  equation~(\ref{covhess}) we did not consider at all the actual value of $\chi^2_m$. We just imposed the conditions \ref{partchi2}, which define an implicit relation of the parameters and the experimental values $\hat y_i$, and carried out uncertainty propagation. Notice, however, that the standard deviations of the parameters $\sigma(a_j)$ scale with 
the experimental uncertainties $\sigma_i$. If the latter are reduced by a factor of two, the former are also reduced by a factor of two. Indeed, if this is done, the value of $\chi^2_m$ will increase by a factor of 4, as will be the Hessian matrix, but this will not modify the values of the best fit $\hat a$.

\subsection{Maximum likelihood method}

We shall describe here the maximum likelihood method, following \cite{cowan98}. The application of the maximum likelihood method requires the
assumption of a The likelihood function ${\cal L}(a)$ as a function of the parameters. If this function is chosen as  ${\cal L}(a) = \exp(-\chi^2(a)/2)$, then the maximum likelihood method gives the same results as the least square method. More generally, one can choose ${\cal L}(a)$ as a decreasing function of $b=\chi^2(a)$. This ensures that the maximum likelihood occurs when $\chi^2(a) = \chi^2_m $, which leads to $a = \hat a$, coinciding with the least square value.   At this stage we will not enter in the specific evaluation of $b$, which will be considered as a parameter. 


Let us consider the region $  \Gamma[b]   $ in the M-dimensional  parameter space that fulfill $b \ge \chi^2(a)  \ge \chi^2_m$. 
The volume in parameter space of this region is
\begin{equation}
V(b) = \int_{\Gamma[b]} P_0(a)\;  d^M \!a = \int   d^M\! a \;  P_0(a) H(b - \chi^2(a)),
\end{equation}
Where $H(x)$ is the Heaviside function, and $P_0(a)$ is an arbitrary function, that defines a metric in the parameter space. It will be later identified with the prior probability distribution in the Bayesian approach.  
We can now consider the region  $\Gamma[b; db]$ which the boundary of the region $  \Gamma[b]   $, with a small thickness $db$, This contains the points that  fulfill  $b \le \chi^2(a) \le b + db$.
The volume of the region $\Gamma[b; db]$, can be expressed as $db \; S(b)$,  where $S(b)$ can be visualized  as the surface of   $  \Gamma[b]   $,  is the derivative of the volume $V(b)$ with respect to $b$
\begin{equation}
S(b) = {d V(b) \over db} = \int   d^M\! a \;   P_0(a) \delta(b - \chi^2(a)) \label{sb}
\end{equation}
The parameter values which are in the region $\Gamma[b, db]$, for small $db$, are described by a covariance matrix which is given by

\begin{multline}
\text{cov}[b](a_j, a_k) =  { 1 \over S(b)} \int d^M\!a \;  P_0(a) \delta(b - \chi^2(a)) \\
\times (a_j - \hat a_j)(a_k - \hat a_k) \label{covb}
\end{multline}

This magnitude can be interpreted as the geometrical average of $(a_j-\hat a_j) (a_k-\hat a_k)$ evaluated in the region $\Gamma[b; db]$, characterized by  $\chi^2(a) \simeq b $.

For a practical evaluation of $   cov[b](a_j, a_k)  $ we can take two approaches, depending on the expected behavior of the function $\chi^2(a)$.

Simplified approach: Let is express $\chi^2(a)$ in the vicinity of the minimum $\hat a$, assuming a quadratic approximation of the $\chi^2(a)  $ function. Its values are then determined by the Hessian function
\begin{equation}
  \chi^2(a) \simeq \chi^2_m + {1 \over 2} \sum_{kj} H_{kj}(a_j-\hat a_j) (a_k-\hat a_k)   \label{chi2quadratic}
\end{equation}
Let us consider the values of the parameters $a_j$ in the region $\Gamma[b; db]$, They define an M-dimensional ellipsoid defined as
\begin{equation}
{1 \over 2} \sum_{kj} H_{kj} (a_j-\hat a_j) (a_k-\hat a_k)  = (b-\chi^2_m) \label{ellipsoid}
\end{equation}

To do the integral, we perform a change of variables so that we go from the $M$ variables $(a_j- a_j^m)$ which fulfill the quadratic equation (\ref{ellipsoid}), to $M$ variables $\xi_l$ which fulfill $\sum_i \xi_l^2 = 1$. After some algebra, following \cite{cowan98} the integral becomes 
\begin{equation}
   cov[b](a_j, a_k)   = {b - \chi^2_m \over M} 2  H^{-1}_{kj} = f[b]^2 cov_{LS}(a_j, a_k)   \label{covbsmooth}
\end{equation}
where $f[b] = \sqrt{(b-\chi^2_m)/M}$ is an enhancement factor which  describes the geometrical average in terms of the parameter $b$ which describes the ellipsoid. 
We see that the geometric covariance is proportional to the least square covariance, provided that the $\chi^2(a)$ surface is parabolic. Besides, when $b = \chi_m^2 + M$, the geometric covariance coincides with the least square covariance.  It should be noticed that often (i.e. \cite{cowan98}) the parameter uncertainties are graphically related to the tangent of the ellipsoid $\chi^2(a)= \chi^2_m + 1$. Here the geometric covariance is not calculated as the tangent, but as the integral previously defined over all the  points of the ellipsoid defined by $\chi^2(a)= \chi^2_m + M$.  The uncertainties of the parameters $\sigma(a_k)$ so obtained coincide with the tangent of the ellipsoid $\chi^2(a)= \chi^2_m + 1$. 

General approach: We have a $\chi^2(a)$ surface which is not parabolic in the region $  \chi^2(a) \simeq b $. We do not even have any analytic expression for $\chi^2(a)$. We can then consider the case in which we sample the parameter space, considering a series of sampling points $a^\ell$ which cover
the parameter space. Each one covers a volume in parameter space denoted by $\Delta V_\ell$, which is the integral of $P_0(a)$ extended to the relevant ranges $\Delta a_j$ of the $M$ parameters. These points have values $\chi^2_\ell$.  In order to perform the integral in equation (\ref{sb}), we substitute the $\delta(b- \chi^2(a))$ function by a more smooth peaked function, which can cover the sampling points, for example a gaussian $\sqrt{c \over \pi}\exp(- c( \chi^2_\ell - b)^2)$. The approximate result of the integral becomes

\begin{equation}
 S(b) \simeq \sum_\ell  {\Delta V_\ell} \sqrt{c \over \pi}  \exp(- c( \chi^2_{\ell} - b)^2)  \label{Sapprox}
 \end{equation}
 This expression allows to define normalized weights $w_\ell(b)$  given by

\begin{eqnarray}
    w_\ell(b) &=& 
\frac{ \exp\big( - c(\chi^2_\ell - b)^2 \big) \, \Delta V_\ell }
     { \sum_{\ell'} \Delta V_{\ell} \exp\big( - c(\chi^2_{\ell'} - b)^2 \big) } \nonumber \\
&=& \frac{ \exp\big( - c(\chi^2_\ell - b)^2 \big) \, \Delta V_\ell }
       { S(b) \sqrt{\pi / c} }
\label{welb}
\end{eqnarray}

\noindent $c$ is a convergence parameter, of the order of 1, which should be chosen large enough to produce a peaked gaussian, but small enough to allow for sufficient sampling points to participate in the expression (\ref{Sapprox}).  A reasonable choice of $c$, should produce smooth, c-independent surface function $S(b)$ and weights  $w_\ell(b)$, for $\chi^2_m \le  b  \le   \chi^2[3 \sigma]$. Once $c$ is converged, and $ w_\ell(b)  $ are determined, one gets
\begin{equation}
cov[b](a_j, a_k) \simeq  \sum_\ell  w_\ell(b)  \; (a^\ell_j-\hat a_j) (a_k^\ell-\hat a_k)  \label{covbgeneral}
\end{equation}

Note that in the simplified approach, as the maximum likelihood covariance matrix is proportional to the least square covariance matrix, the two correlation matrices are  identical. However, in the general approach the maximum likelihood correlation matrix may differ significantly from the least square one. 

\subsection{Bayesian approach}

 Bayes theorem states
 \begin{equation}
P(A|H) = {P(H|A) \over P(H)} P(A)  \label{bayes}
\end{equation}
Let us specify this expression for our case. $P(A)$ is the probability that the model that we use has parameters $a$ are in a certain region $A$ of the parameter space. This probability is previous to consideration of the experimental data, so this is the prior probability. $P(A|H)$ is the probability that the model that we use has parameters are in a certain region, taking into account the experimental data. This is the posterior probability, which is what we want to determine. Note that the p-value that characterizes the region $A$ is precisely $1-P(A|H)$.
$P(H|A)$ is the probability of occurrence of the experimental data, given the parameter region considered. This will be given in terms of the $\chi^2$ probability distribution, where $\chi^2$ determines the discrepancy of the experimental data denoted by $H$ and the calculations of the parameters included in $A$. Finally, $P(H)$ is the probability of occurrence of the experimental data, for any possible value of the parameters, and it is  obtained from the normalization of  $P(A|H)$.

If the parameter region is chosen as $A=\Gamma[b]$, which is characterized by $b \ge\chi^2(a) \ge \chi^2_m $, then the conditional probability $P(H|\Gamma[b])$ is given by the complementary of the cumulative $\chi^2$ probability distribution, $1-P(b, L)$.  Note that $P(b,L)$ was determined as the probability that the randomly generated $\chi^2(r)$ values, consistent with the data probability distribution,  are larger than the reference value $\chi^2=b$.  $P(H|\Gamma[b])$ is the probability that they are smaller, so  $P(H|\Gamma[b]) = 1- P(b,L)$. Thus, when $b=0$, $P(H|\Gamma[b]) = 0$ and when $b \to \infty$, $P(H|\Gamma[b]) \to 1$.
The conditional probability is additive, so that if the parameter region $\Gamma[b]$ is divided into several disjoint regions, then $P(H|\Gamma[b])$ will be the sum of the conditional probabilities of the different regions. In particular, If the parameter region is chosen as the narrow boundary $\Gamma[b;\Delta b]$, which is characterized by $b\ge\chi^2(a) \ge b - \Delta b$, then $P(H|\Gamma[b;\Delta b])$is given by the differential $\chi^2$ probability distribution $-{dP(b, L) \over db} \Delta b$. As $P(b,L)$ decreases with b,  $P(H|\Gamma[b;db])$ is positive, as it should.
Note that the volume in parameter space of the region $\Gamma[b;\Delta b]$ is $S(b) \; \Delta b$.  The region
$\Gamma[b;\Delta b]$ can be divided into several small disjoint regions in parameter space $\Gamma[a;\Delta^M \!a]$, each of them which has a volume in parameter space  $\Delta^M\! a $, around a parameter vector $a$. The conditional probability  $P(H| \Gamma[a;\Delta^M\! a]  )$ is proportional to the volume in parameter space $\Delta^M\! a$, so we obtain
\begin{eqnarray}
      P(H|\Gamma[b]) &=&  1-  P(b, L) \\
      P(H|\Gamma[b;\Delta b]) &=&  - {dP(b, L) \over db} \Delta b \\
      P(H| \Gamma[a;\Delta^M a]  ) &=&   -{dP(b, L) \over db} { \Delta^M\! a \over S(b)}
\end{eqnarray}

From these expressions we can apply Bayes theorem. The prior probability $P(A)$ is obtained integrating a prior probability density ${\cal P}_0(a) $ over a certain parameter region $\Gamma$ . The posterior probability density $P(A|H) $ is obtained  integrating the posterior  probability density ${\cal P}_B(a) $ which we want to obtain. The relation of both is just
\begin{equation}
    {\cal P}_B(a) =   {-1 \over P(H)}{dP(b, L) \over db} { 1 \over S(b)}{\cal P}_0(a) 
\end{equation}
where $b= \chi^2(a)$. This expression can be more conveniently written as 
\begin{equation}
    {\cal P}_B(a) =   {-1 \over P(H)} \int_{\chi^2_m}^\infty db {dP(b, L) \over db} { \delta(b-\chi^2(a)) \over S(b)}{\cal P}_0(a) 
\end{equation}
The value of $P(H)$ is obtained from the normalization condition 
\begin{eqnarray}
  1 &=&  \int d^M\! a  \;  {\cal P}(a) \\
  &=&  {-1 \over P(H)} \int_{\chi^2_m}^\infty db {dP(b, L) \over db} \int d^M\! a  \; { \delta(b-\chi^2(a)) \over S(b)} {\cal P}_0(a) \nonumber
\end{eqnarray}
The definition of $S(b)$, eq.\ref{sb} makes the last integral equals to 1. Thus
\begin{equation}
1 = {-1 \over P(H) }  \int_{\chi^2_m}^\infty  db \; {dP(b, L) \over db}    = {1 \over P(H) }  P(\chi^2_m, L).
\end{equation}
From this, the explicit expression for the unconditional data probability $P(H)=    P(\chi^2_m, L)  $ is obtained, and the posterior probability distribution becomes
\begin{equation}
    {\cal P}_B(a) =   {-1 \over P(\chi^2_m, L)}  {dP(b, L) \over db} { 1 \over S(b)} {\cal P}_0(a) 
\end{equation}

Note that the effect of proposing different values for the prior probability distribution $ {\cal P}_0(a)  $ only affects the distribution of the probabilities for the different parameter values that fulfill $b \ge \chi^2(a) \ge b- \Delta b$. The total posterior probability of the region $\Gamma(b, Delta b]$ is
\begin{equation}
        P(\Gamma[b;\Delta b]|H) =  - {dP(b, L) \over db} {\Delta b \over   P(\chi^2_m, L)} \\
\end{equation}
and becomes independent of the prior distribution.

Indeed, if one makes use of a prior probability distribution $ {\cal P}_0(a)  $ which depends strongly on $a$, then the contours in the posterior probability distribution $ {\cal P}_B(a)  $ will be distorted, and will no longer be determined by the value of $b=\chi^2(a)$. That does not alter the posterior probabilities of the region $  \Gamma[b;\Delta b] $, but would indicate that the proper construction of the region in parameter space  associated to a given p-value should be associated to contours of  $ {\cal P}_B(a)  $, which are affected by  $ {\cal P}_0(a)  $.   In this paper, we will consider,  in agreement with refs \cite{Nunes2017, Nunes2019}, that the analysis is data-driven, so the distortion in the p-value regions associated to $ {\cal P}_0(a)  $ can be neglected.



From the parameter distribution ${\cal P}_B(a)$, we want to evaluate distribution properties, in particular the covariance matrix. Notice that the magnitude that we want to evaluate is different (although related to) the geometric covariance matrix    $cov[b](a_k, a_j)$
or the least square covariance matrix  $cov_{LS}(a_k, a_j)$.   The quantity that we want to evaluate is
\begin{equation}
cov_B(a_j, a_k) =  \int d^M\!a \; {\cal P}_B(a) (a_k - \hat a_k) (a_j - \hat a_j)
\end{equation}
Using the previous expressiones, we have that the covariance for the probability distribution ${\cal P}(a) $ can be written in terms of the covariances for the geometrical distributions as 
\begin{equation}
cov_B(a_j, a_k) =   \int_{\chi^2_m}^\infty d b {d P_r(b) \over db} cov[b  ](a_j, a_k) \label{covbayesian}
\end{equation}
From this expression, we can proceed, differenciating the two cases for the evaluation of $cov[b  ](a_j, a_k)$, which are the simplified expression \ref{covbsmooth} and the general expression \ref{covbgeneral}


Using the simplified expression \ref{covbsmooth}, and implementing analytical properties of the $\chi^2$ distribution functions, we get
\begin{eqnarray}
cov_B(a_j, a_k) &=&  F^2_B \;   cov_{LS}(a_j, a_k) \label{covbayesiansmooth1}\\
F^2_B  &=&   {L \; P(\chi^2_m, L+2) - \chi^2_m \; P(\chi_m^2, L) \over M \; P(\chi_m^2, L)}
\end{eqnarray}

This expression indicates that the Bayesian approach produces a systematic enhancement of the covariance matrices, as compared to the least square approach.  This enhancement  is typically about $\sqrt{L}/M$, for good fits for which $\chi^2_m \simeq L$. It gets smaller for poorer fits, $\chi^2_m > L$, and larger for fits that are too good $   \chi^2_m < L $. In Table \ref{tablechi2} we present the relevant values for $f[b](\chi_m^2, L)$ and  $F_B(\chi_m^2, L)$ for a case of interest.

Using the general expression \ref{covbgeneral},  we get
\begin{eqnarray}
cov_B(a_j, a_k) &=&  \sum_\ell w_\ell   (a_k^\ell - \hat a_k) (a_j^\ell - \hat a_j) \label{covbayesiangeneral} \\
w_\ell   &=&   \int_{\chi^2_m}^\infty d b {d P_r(b) \over db} w_\ell(b) \nonumber  \\
\label{weightcorrect}
   &\simeq& \left. {d P_r(b) \over db} \right|_{b=\chi^2_\ell} {\Delta V_\ell  \over   S(\chi^2_\ell)} 
\end{eqnarray}

\nocite{*}

\bibliography{apssamp}

\end{document}